\documentclass[10pt,here,feynmf]{article}

\usepackage{feynmf}
\usepackage[all]{xy}
\usepackage{amsmath,amssymb}
\usepackage{graphicx}

\usepackage{epstopdf}

\begin{document}

\begin{center}
\scshape
CENTRE DE PHYSIQUE TH\'EORIQUE \footnote{\, Unit\'e Mixe de
Recherche (UMR) 6207 du CNRS et des Universit\'es Aix-Marseille 1 et 2 \\ \indent \quad \, Sud Toulon-Var, Laboratoire affili\'e \`a la 
FRUMAM (FR 2291)} \\ CNRS--Luminy, Case 907\\ 13288 Marseille Cedex 9\\
FRANCE\\
\bigskip
\centerline{***}
\bigskip
INSTITUT F\"UR THEORETISCHE PHYSIK\\
GEORG-AUGUST UNIVERSIT\"AT G\"OTTINGEN\\
Friedrich-Hund-Platz 1\\
37077 G\"ottingen\\
DEUTSCHLAND

\end{center}
\vspace{1cm}
\begin{center}
{\huge\bfseries Wilsonian renormalization, differential equations and Hopf algebras}
\end{center}

\vspace{1cm}

\begin{center}

{\Large T. Krajewski}\footnote{Centre de Physique Th\'eorique, Marseille, {\tt krajew@cpt.univ-mrs.fr}}, 
\quad
{\Large P. Martinetti}\footnote{Institut f\"ur Theoretische Physik, G\"ottingen, {\tt martinetti@theorie.physik.uni-goettingen.de}}  
\vspace{1cm}

\end{center}

\begin{center}
{\it \large Talk given by T. Krajewski at the conference "Combinatorics and Physics"}\\
\vspace{0.5cm}
{\it \large Max Planck Institut F\"ur Mathematik}\\
\vspace{0.5cm}
{\it \large Bonn, March 2007}\\
\end{center}
\vspace{1cm}

%{\bfseries\scshape Abstract:} 

\subsection*{Abstract}
In this paper, we present an algebraic formalism inspired by Butcher's B-series in numerical analysis and the Connes-Kreimer approach to perturbative renormalization. We first define  power series of non linear operators and propose several applications, among which the perturbative solution of a fixed point equation using the non linear geometric series. Then, following Polchinski, we  show how perturbative renormalization works for a non linear perturbation of a linear differential equation that governs the flow of effective actions. Furthermore, we define a general Hopf algebra of Feynman diagrams adapted to iterations of background field effective action computations.  As a simple combinatorial illustration, we show how these techniques can be used to recover the universality of the Tutte polynomial and its relation to the $q$-state Potts model. As a more sophisticated example, we use ordered diagrams with decorations and external structures to solve the Polchinski's exact renormalization group equation. Finally, we work out an analogous construction for the Schwinger-Dyson equations, which yields a bijection between planar $\phi^{3}$ diagrams and a certain class of decorated rooted trees.

\vspace{0.5cm}

 %\vskip 1truecm
%PACS-93:\\
%\indent MSC-91: 
%\vskip 1truecm
\noindent June 2008\\
\noindent
CPT-P36-2008

\newpage

\tableofcontents

 \section{Introduction}

The last century has witnessed the development of Quantum Field Theory (QFT) as the adequate framework for the formulation of many physical theories, ranging from elementary particles to condensed matter physics. Historically, it was first devised to describe the interaction of charged particles with the electromagnetic field, leading to Quantum Electrodynamics (QED). However, QED was shown to be plagued by a surprising disease: When expanded in powers of the electric charge $e$, corrections to the leading terms turn out to be divergent. The occurrence of these infinities can be traced to virtual processes involving very high energies.  Then, if we limit the energy these processes involve to a scale $\Lambda$ and choose the electric charge $e(\Lambda)$ as a suitable function of $\Lambda$, all the divergent quantities can be made finite, provided they are expressed in terms of a few low energy parameters to be determined by experiment, like the masses of the particles and the strength of their interactions. This is the basic idea of renormalization: Processes occurring at a given scale must be formulated using parameters adapted to these scale.

Later on, QFT progressively invaded almost all fields of theoretical physics.  Renormalizability was a guideline for the construction of the standard model of elementary particles while critical phenomena in condensed matter were also understood using renormalization. More in general, QFT describes fluctuating systems with continuous degrees of freedom and is best understood through paths integrals of the type
\begin{equation}
\int[D\phi]\,\mathrm{e}^{-S[\phi]},
\end{equation}
where the integration is over a space of functions. When considered as a perturbation of a Gau\ss ian  integral, this path integral is expanded over Feynman diagrams. Because of the continuous nature of the path integral, some of these diagrams yield divergent quantities, as dictated by dimensional analysis, and can only acquire a meaning through renormalization. In a certain sense, it can be said that the founding fathers of QFT (Dirac, Feynman,  Schwinger, Tomonoga, Dyson, Weinberg, t'Hooft, Veltman, Wilson and many others) have achieved an infinite dimensional analogue of Newton's work: Driven by fundamental physics, they devised a new form of calculus based on path integrals, Feynman diagrams and renormalization.

However, it is fair to say that this new calculus is not yet fully understood. In particular, it hides an algebraic structure analogous to diffeomorphisms, as uncovered by Connes and Kreimer a decade ago (see \cite{ck0}, \cite{ck1} and \cite{ck2} as well as the first chapter of the book \cite{connesmarcolli} and \cite{kk} for recent reviews). To be specific, divergent Feynman diagrams generate a commutative Hopf algebra whose associated group is a refinement of the group of diffeomorphisms of the coupling constant. This formalism is versatile enough to encompass the infinite dimensional analogue of the change of parameters provided by the computation of the Wilsonian effective action, as we show in this paper. Our contribution is organized as follows.

In the first section, we give a terse account of renormalization. We first recall the elementary particle physics point of view and introduce Feynman diagrams. Then, we introduce Wilson's viewpoint, first for spin systems and then for general QFT. This is textbook physics and can be found in modern texts on QFT (to quote only a few authors who strongly influenced our presentation, see the textbook \cite{zee}  by Zee for an eclectic overview of QFT applied to particle and condensed matter physics, the treatise by Zinn-Justin \cite{zinn} for a general treatment of QFT in the path integral formalism  and the review by Rivasseau \cite{Rivasseau} for an introduction to renormalization, accessible to a large audience). We close this section by recalling a renormalization group proof of the central limit theorem of probability theory that may help mathematicians grasp Wilson's ideas.

The next section is devoted to rooted trees and their Hopf algebra. These were introduced as fundamental tools in the numerical analysis of a differential equation by Butcher \cite{butcher}. We revisit this construction and interpret trees as indices for power series of non linear operators, with composition law given by the convolution product. We illustrate this construction with the non linear geometric series that provides a pertubative solution of a general fixed point equation. This can be used to resum tree like structures as shown on a elementary example based on Gau\ss ian random matrices.

In the last section, we deal with renormalization. We first formulate perturbative renormalization for a general renormalization group differential equation. Then, we introduce a general Hopf algebra of Feynman diagrams that encodes successive effective action computations. We illustrate the use of this Hopf algebra on two examples. First, we give a new proof of the universality of Tutte polyomial and of its relation to the Potts model. Then, we use the Hopf algebra to solve Polchinski's equation that describes the flow of effective actions in QFT. Finally, in the last section we investigate an analogous structure based on the Schwinger-Dyson equations. As a byproduct, we obtain a new bijection between planar $\phi^{3}$ diagrams and a certain class of decorated rooted trees.

\section{Basics of wilsonian renormalization}

\subsection{What is renormalization?}

In its most general acceptance, renormalization can be defined as the general methods that implement the change of parameters in the description of a physical system necessary in order to take into account the indirect effects of unobserved degrees of freedom, usually living on a much shorter length scale. The simplest example we can think of is the propagation of light in a medium, as described by geometrical optics. Whereas in the vacuum the speed of light is $c$, in some medium it is reduced to $c/n$, with $n$ the refractive index, as a consequence of the interaction of light with the atoms of the medium. While the latter are not described by geometrical optics, their effect on the propagation of light is simply taken into account by the substitution $c\rightarrow c/n$.

As a second example, let us consider the motion of a ping-pong ball of inertial mass $m_{\mathrm{inert}}$, gravitational mass $m_{\mathrm{grav}}$ and volume $V$ in a perfect fluid of density $\rho$, as popularized by Connes \cite{connesmarcolli}. A straightforward application of Newton's law of dynamics yields its acceleration through 
\begin{equation} 
m_{\mathrm{inert}}{\bf a}=m_{\mathrm{grav}}{\bf g}-\rho V{\bf g},
\end{equation}
where the first term is the ball's weight and the second one Archimedes' buoyant force. If the ball is ten time lighter than the equivalent volume of water, then it experiences an acceleration ${\bf a}=-9\,{\bf g}$, which is obviously nonsensical. The hole in the previous reasoning is that we did not take into account the inertia of the surrounding fluid. Using a suitable ansatz for the fluid's velocity field, it can be shown \cite{connesmarcolli} that the latter implies a shift of the inertial mass of the ball 
\begin{equation}
m_{\mathrm{inert}}\rightarrow m_{\mathrm{inert}}+\frac{\rho V}{2},
\end{equation}
which yields the more acceptable acceleration of ${\bf a}=-3/2\,{\bf g}$.  Both these elementary examples illustrate the main feature of renormalization: In a simple theory (geometrical optics, Newton's law of dynamics), the effect of unobserved degrees of freedom, which should be described by a more sophisticated theory (electrodynamics, fluid mechanics), are taken into account by a mere change of parameters in the simpler theory.

The very same type of ideas show off in QFT, which grew up from the marriage between special relativity and quantum mechanics. First, recall that one of the main lessons from special relativity is the equivalence between matter and energy, as neatly summarized by Einstein's celebrated relation $E=mc^{2}$. This is currently observed in high energy physics experiments: Incoming particles (say leptons, like a pair $e^{+}e^{-}$) with a total given energy collide and result in some outgoing particles whose nature may be totally different (for instance, a pair $q\overline{q}$).  Of course, total energy is always conserved so that heavy particles may appear only if the incoming energy is high enough.
     
On the other hand, quantum mechanics provides us with a description of the microscopic world. Its predictions, carefully verified by experiments, yield the probability of finding the system in a final state $|f\rangle$ knowing that it was initially in the state $|i\rangle$. The transition amplitude from an initial state to the final one is often expressed as a sum over all (unobserved) intermediate states $|n\rangle$,
\begin{equation}
\langle f|i\rangle=\sum_{n}\langle f|n\rangle\langle n|i\rangle.
\end{equation}
In QFT, these initial and final states represent a fixed number of elementary particles of given energy but the summation over all the intermediate states involves an arbitrary number of virtual particles of arbitrarily high energies. It is fundamental to note that none of the intermediate states are observed since the experiment only answers the question \emph{"What is the probability of finding the system in the state $|f\rangle $ at  the end of the experiment if we know that it was initially in the state $|i\rangle$?"}.  

In most cases of interest, the sum over intermediate states diverges as the upper limit of the energy $\Lambda$ of the virtual particles go to infinity. However, for the so-called renormalizable theories, it turns out that these infinities cancel  when the parameters of the theory (masses $m(\Lambda)$ and coupling constants  $g(\Lambda)$) are suitably chosen as functions of  $\Lambda$, which may then be taken to infinity leaving physically relevant quantities finite. This is the process of renormalization, as it occurs in high energy physics: The effect of unobserved virtual states of high energy is taken into account in some of the parameters of the theory. In a certain sense, the cut-off $\Lambda$ sets a limit for the validity of our field theory  and the sole effect of the very high energy modes beyond $\Lambda$ is to adapt the parameters of the theory to energy of the process we consider.  Alternatively, this can also be considered as an ignorance of the very small scale structure of space-time since high energy correspond to small distances by Fourier transform.

\subsection{Feynman diagrams and perturbative renomalization}

As far as renormalization is concerned, QFT is best studied in the framework of the Euclidian path integral. In the simplest case, the fields are defined as functions $\phi$ from space-time ${\Bbb R}^{D}$ to ${\Bbb R}$ and the dynamics is governed by the action functional,
\begin{equation}
S[ \phi]=\int d^{D}x
\left(\frac{1}{2}\partial_{\mu}\phi\,\partial^{\,\mu}\phi+
\frac{1}{2}{m}^{2}\phi^{2}+
\frac{g}{N!}\phi^{N}\right),
\end{equation} 
where $m$ is a mass and $g$ a coupling constant for a $N$-particle interaction. At the quantum level, all the information is encoded in the Green's functions,
\begin{equation}
G(x_{1},\dots,x_{n},m,g)={\cal N}
\int[D\phi]\,\mathrm{e}^{-\frac{\cal{S}[ \phi]}{\hbar}}\phi(x_{1})\cdots\phi(x_{n}),\label{green}
\end{equation}
where the integration is over the space of all fields and ${}\cal N$ is a normalization constant to be defined later.  $\hbar=6.02\,10^{-34}$ J$\cdot$s is Planck's constant and measures the deviation of the quantum theory from the classical one. In most of the paper, we choose unit such that $\hbar=1$. Note that is is only the simplest model of a scalar field theory. More complicated action functionals are required to account for the real world physics: Spinors $\psi$ for fermionic particles and gauge connections $A$ for their interactions, as is the case for QED and the Standard Model of elementary particles. Nevertheless, we restrict our attention in the sequel to the simplest example of a scalar field theory.  

To give a precise meaning to \eqref{green}, it is convenient to expand  $G(x_{1},\dots,x_{n},m,g)$ as a power series in $g$ using Feynman diagrams. The latter are best introduced on simpler analogue, with the space of fields replaced by a finite dimensional vector space. Thus, the equivalent of \eqref{green} is
\begin{equation}
G_{i_{1},\dots,i_{n}}(A,V)={\cal N}\int d\phi\,\mathrm{e}^{-\cal{S}(\phi)}\phi_{i_{1}}\cdots\phi_{i_{n}},
\label{greenfinite}
\end{equation}
with the action
\begin{equation}
S(\phi)=\frac{1}{2}\phi\cdot A^{-1}\cdot\phi +V(\phi).
\end{equation}
The quadratic term is defined by a positive definite symmetric matrix $A$,
\begin{equation}
\phi\cdot A^{-1}\!\cdot\phi=\sum_{i,j}(A^{-1})_{ij}\phi_{j}\phi_{i},
\end{equation}
and the interaction potential $V$ is a polynomial in all the fields
\begin{equation}
V(\phi)=\sum_{N}\sum_{i_{1},\dots,i_{N}}\frac{g_{i_{1},\dots,i_{N}}}{N!}\phi_{i_{1}}\cdots\phi_{i_{N}}.
\label{finitepot}
\end{equation}
In this case we choose ${\cal N}^{-1}=\det A/2\pi$ and expand $\mathrm{e}^{-V(\phi)}$ as a power series in the couplings $g_{i_{1},\dots,i_{N}}$. Thus, \eqref{greenfinite} amounts to the computation of the average of a monomial using a Gau\ss ian weight and is given by Wick's theorem: It is a sum over all possible pairings of the variables in the monomial, each pairing of $\phi_{i}$ with $\phi_{j}$ being weighted by $A_{ij}$. For example,
 \begin{equation}
 \phi_{i} \phi_{j} \phi_{k} \phi_{l}\quad\rightarrow\quad
A_{ij}A_{kl}+A_{ik}A_{jl}+A_{il}A_{jk}.
\end{equation}
Then, each term of the expansion of \eqref{greenfinite} is associated with a diagram with $n$ external legs and vertices of valence $N$,
\begin{equation}
G_{i_{1},\dots,i_{n}}(A,V)=\sum_{\gamma}\frac{G^{\,\gamma}_{i_{1},\dots,i_{n}}(A,V)}{\mathrm{S}_{\gamma}}\label{finitediag}.
\end{equation}
The contribution of each diagram is computed using the Feynman rules:
\begin{itemize}
\item
associate the indices $i_{1},\dots,i_{n}$ to the external legs and indices $j_{k}$ to the internal half edges;
\item
associate a matrix element $A_{j_{k}j_{l}}$ to any edge connecting the indices $j_k$ and $j_l$;
\item
associate a coupling $-g_{j_{1},\dots,j_{N}}$ to a $N$-valent vertex  whose half edges have indices $j_{1},\dots,j_{N}$;
\item
sum over all the indices $j_{k}$.
\end{itemize}
Besides, on has to divide by the symmetry factor $\mathrm{S}_{\gamma}$ which is the cardinal of the automorphism group of the diagram, leaving the external legs fixed. For example,
\begin{equation}
i_{1}\,
\xy
(0,0)*{}="A"; 
(16,0)*{}="B"; 
(-4,0)*{}="C"; 
(20,0)*{}="D"; 
"C"; "A" **\crv{(-4,0)&(0,0)};
"A"; "B" **\crv{(0,0)&(8,8)&(16,0)};
%"A"; "B" **\crv{(0,0)&(10,0)};
"A"; "B" **\crv{(0,0)&(8,-8)&(16,0)};
"B"; "D" **\crv{(16,0)&(20,0)};
\endxy
\,i_{2}\quad
\rightarrow\quad
\frac{1}{2}
\sum_{j_{1},j_{2},j_{3}\atop
j_{4},j_{5},j_{6}}
A_{i_{1},j_{1}}
g_{j_{1},j_{2},j_{3}}
A_{j_{2},\,j_{4}}A_{j_{3},\,j_{5}}\,
g_{j_{4},j_{5},j_{6}}
A_{j_{6},\,i_{2}}.
\end{equation}
This simple finite dimensional model already captures  some important algebraic aspects of perturbation theory as will be discussed in the last part devoted to the Hopf algebras based on Feynman diagrams.

At a formal level, the Green's functions \eqref{green} can be computed as a power series in $g$ by replacing $\phi_{i}$ by a function $x\mapsto\phi(x)$, the matrix element $A_{ij}$ by the propagator
\begin{equation} 
K(x,y)=\int_{{\Bbb R}^{D}}\frac{d^{D}p}{(2\pi)^{D}}\,\frac{\mathrm{e}^{\mathrm{i}p\cdot(x-y)}}{p^{2}+m^{2}}.
\end{equation}
and $V(\phi)$ by the interaction term
\begin{equation}
\frac{g}{N!}\int_{{\Bbb R}^{D}} d^{D}x\,\phi^{N}(x)=
\frac{g}{N!}\int_{{\Bbb R}^{D}}\frac{d^{D}p_{1}}{(2\pi)^{D}}\cdots\frac{d^{D}p_{N}}{(2\pi)^{D}}\,(2\pi)^{D}\delta(p_{1}+\cdots+p_{N})\tilde{\phi}(p_{1})\dots\tilde{\phi}(p_{N}),
\end{equation}
with 
\begin{equation}
\tilde{\phi}(p)=\int_{{\Bbb R}^{D}}d^{D}x\,\mathrm{e}^{-\mathrm{i}p\cdot x}\,\phi(x)
\end{equation}
the Fourier transform of $\phi(x)$. Heuristically, the Feynman diagrams can be thought of as quantum mechanical processes with particles on their external legs and virtual particles of momenta $p$ propagating on the internal lines. It is important to notice that although momentum is conserved at each vertex and along each line, the particles that propagate along the loops\footnote{What is called loop in the physics terminology adopted here is called a cycle by graph theorists.} may have arbitrary momenta.

The integral along loop momenta extend on the whole of ${{\Bbb R}^{D}}$ and may led to divergencies since the Fourier transform of propagator does not decrease fast enough at large $p$. A connected diagram with $I$ internal lines, $V$ vertices and $L=I-V+1$ loops behaves  in dimension $D$ like $p^{\omega}$ with $\omega=LD-2I$ the superficial degree of divergence. Using $NV=2I+E$, the latter can be written as 
\begin{equation}
\omega=V\left(\frac{ND}{2}-D-N\right)+E\left(1-\frac{D}{2}\right)+D
\end{equation}
When the coefficient of $V$ is strictly positive, divergencies occur for any Green's function and such a theory cannot be renormalized. If this coeffcient is strictly negative, then there are only a finite number of divergent diagrams. These are the superrenormalizable theories which are nevertheless of limited interest in physics. Finally, the critical situation defines the renormalizable theories, like $\phi^{4}$ in $D=4$ or $\phi^{3}$ in $D=6$. In this case, the divergencies occur only for the diagrams with $E\leq N$ and may absorbed into the coefficients of a polynomial  interaction of degree $N$. At the notable exception of gravity, all the interactions of elementary particles can be formulated using renormalizable interactions. 

To prevent the propagation of Fourier modes of momenta $\geq \Lambda$, let us alter the propagator by introducing a cut-off $\Lambda$,
\begin{equation} 
K(x,y)\,\rightarrow\,K_{\Lambda}(x,y)=\int_{\frac{1}{\Lambda^{2}}}^{\infty} d\alpha\,\int_{{\Bbb R}^{D}}\frac{d^{D}p}{(2\pi)^{D}}\,
\mathrm{e}^{\mathrm{i}p\cdot(x-y)}
\mathrm{e}^{-\alpha(q^{2}+m^{2})}
\label{propagator}.
\end{equation}
This procedure is known as the regularization and can be performed in various ways: Besides  the method used here, one could also discretize the theory on a lattice or evaluate the diagrams in complex dimension $z$ and recover the divergences as poles when $z\rightarrow D$. In principle all these methods are equivalent but we restrict ourselves to the momentum space  cut-off presented here since it is suited to the Wilsonian point of view we adopt in this paper.

For a renormalizable theory like the $\phi^{4}$ theory, one can trade the parameters $g$ and $m$ for some cut-offs dependent ones $g_{0}(\Lambda)$ and $m_{0}(\Lambda)$ and further introduce an additional wave function renormalization $Z(\Lambda)$ in such a way that $Z^{\frac{n}{2}}(\Lambda)G_{\Lambda}(x_{1},\dots,x_{n},m_{0}(\Lambda), g_{0}(\Lambda))$ admits a finite limit when $\Lambda\rightarrow\infty$. To obtain definite physical predictions, the bare parameters $g_{0}(\Lambda)$ and $m_{0}(\Lambda)$ and the wave function renormalization $Z(\Lambda)$ must be determined in terms of  normalization conditions involving renormalized parameters $m_{\mathrm{r}}$ and $g_{\mathrm{r}}$ measured at a low energy scale $\mu$. Thus, we define the renormalized Green's functions as
\begin{equation}
G_{\mathrm{r}}(x_{1},\dots,x_{n},m_{\mathrm{r}},\mu, g_{\mathrm{r}})=\lim_{\Lambda\rightarrow\infty}Z^{\frac{n}{2}}(\Lambda,m_{\mathrm{r}}, g_{\mathrm{r}},\mu)G_{\Lambda}(x_{1},\dots,x_{n},m_{0}(\Lambda,m_{\mathrm{r}}, g_{\mathrm{r}},\mu), g_{0}(\Lambda,m_{\mathrm{r}}, g_{\mathrm{r}},\mu))
\end{equation}
Note that we are dealing here with perturbative renormalization only, so that the previous equality must be understood as an equality between formal power series in $g_{\mathrm{r}}$. In fact, $g_{0}(\Lambda)$, $m_{0}(\Lambda)$ and $Z(\Lambda)$ are themselves formal power series in $g_{\mathrm{r}}$ that can be computed in terms of Feynman diagrams using the Bogoliubov-Parasiuk-Hepp-Zimermann (BPHZ) formula. Roughly speaking, the contribution of a divergent diagram with 2 or 4 external legs to the renormalisation of the parameters is encoded in its counterterm $C(\gamma)$ which is determined recursively by the relation
\begin{equation} 
C(\gamma)=T\Big(\sum_{\left\{\gamma_{i},\dots\right\}\atop \gamma_{i}\cap\gamma_{j}=\emptyset}\prod_{i}C(\gamma_{i})
\frac{\gamma}{\prod_{i}\gamma_{i}}\Big),
\end{equation}
with $T$ taking the divergent part of the diagram. This sum runs over all sets , including the empty one, of disjoint, divergent, one particle irreducible subdiagrams of $\gamma$ (i.e. diagrams that cannot be disconnected by cutting an arbitrary internal line). The reduced diagram on the RHS is obtained by shrinking each $\gamma_{i}$ to a single vertex and finally take the divergent part of the whole sum, with a finite part determined by the normalization conditions at the low energy scale $\mu$. In the framework of dimensional regularization, this operation is elegantly written as a Birkhoff decomposition for a loop in the space complex dimension with values in a group associated to a commutative Hopf  algebra (see 
the work of Connes and Kreimer \cite{ck1} and \cite{ck2}).

\begin{figure}
\begin{center}
\includegraphics[width=3cm]{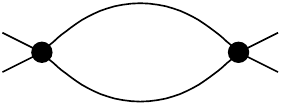}
\caption{One loop diagram for the four point function}
\label{simpleoneloop}
\end{center}
\end{figure}

In the construction of the renormalized theory, the renormalized coupling constant $g_{\mathrm{r}}$ and mass $m_{\mathrm{r}}$ are the parameters of the theory which are directly accessible. They are determined by experiments performed at a low energy scale $\mu$. However, if we consider the same theory but with renormalized parameters determined at a slightly different scale $\mu'$, then we are performing a mere change in the parametrization of the same theory. Consequently, the renormalized parameters measured at different energy scale are not independent. For instance, the renormalized coupling  constant depends on the scale through a differential equation of the type
\begin{equation}
\mu\frac{dg}{d\mu}=\beta(g).
\end{equation}    
The $\beta$ function can be determined by requiring that the bare paramaters, which are the "true" parameters of the theory entering in the path integral, are independent of $\mu$.  Accordingly, the low energy renormalized theory has a subtle form of scale invariance inherited from the behavior of the high energy virtual processes. The zeroes of $\beta$ define the fixed points of the renormalization group  and play a fundamental role. In particular, the origin $g=0$ is a fixed point and it is (UV) attractive if $\beta(g)<0$. This last case corresponds to the asymptotically free theories for which the coupling constant goes to $0$ at high energies.  These theories are believed to be consistent non pertubatively, that is,  the expansion of the Green's functions can be given a meaning beyond that of an asymptotic series at $g=0$. In the other case, there are strong arguments against a non perturbative definition.

For instance, at one loop level for the $\phi^{4}$ model in dimension 4, the relation between the bare and the renormalized theory follows from the computation of the diagram given in figure \ref{simpleoneloop},
\begin{equation}
-g_{\mathrm{r}}=-g_{0}+\beta g_{0}^{2}\log\frac{\Lambda}{\mu}+O(g_{0}^{2}),\label{betaoneloop}
\end{equation}
with $\beta$ a positive constant. Here, $g_{r}$ is identified with the four point function evaluated on a specific configuration of momenta of order $\mu$, using the bare theory with a cut-off $\Lambda$.

\begin{figure}
\begin{center}
\includegraphics[width=10cm]{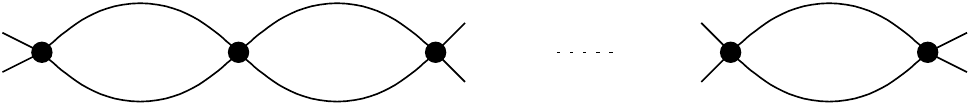}
\caption{Chain of one loop diagrams for the four point function}
\label{chaindia}
\end{center}
\end{figure}

From this relation, one readily extracts the $\beta$ function as $\beta(g)=\beta g^{2}$, so that the theory is not asymptotically free. If we now attempt to get a non perturbative information from \eqref{betaoneloop} by summing the geometric series associated to the chain of one loop diagrams depicted on figure \ref{chaindia} we get, as $g_{0}>0$ for the consistency of the path integral,
\begin{equation}
g_{\mathrm{r}}=\frac{g_{0}}{1+\beta g_{0}\log\frac{\Lambda}{\mu}}\leq\frac{1}{\beta\log\frac{\Lambda}{\mu}},
\end{equation}
so that the $g_{\mathrm{r}}\rightarrow 0$ as $\Lambda\rightarrow\infty$. This is the triviality problem of $\phi^{4}$ theory. Note that the argument would not apply if $\beta$ were negative.

To close this preliminary section on renormalization, let us give a brief historical sketch to show how these ideas gradually appeared in physics.

\begin{itemize}
\item
\underline{\bf Early computations in QED}

It was already noticed by the founding fathers of QED in the thirties that the first non trivial terms of the perturbative expansion are plagued by infinities. However, certain combination of divergent quantities were shown to be finite, thus leading to successfully tested predictions. For instance, the anomalous magnetic moment of the electron, which is nowadays one of the most precisely tested prediction of physics. This set the stage for the general proof of the renormalizabilty of QED in the fifties by  Dyson, Feynman, Schwinger and Tomonoga.

\item
\underline{\bf Perturbative QED  and its $\beta$ function}

For QED with massless fermions an altered form of scale invariance was discovered: If we measure the low energy electric charge $e$ at various energies $\mu$, we describe the same physics provided we allow for a $\mu$ dependent charge $e_{\mu}$. QED has a positive $\beta$ function so that the electric charge increases at high energy. This also indicates a breakdown of (perturbative) QED at very high energy, which is no surprise since QED is not a complete theory as it does not take into account nuclear forces.

\item
\underline{\bf BPHZ (Bogoliubov-Parasiuk-Hepp-Zimmermann) formula}
 
Systematic  combinatorial rules for renormalization were found in the sixities allowing to express the counterterm of a diagrams in terms of counterterms of  its subdiagrams and divergent integrals. This is the content of the BPHZ formula and solved the problem of overlapping divergences.

\item
\underline{\bf Electroweak interactions}

In the late sixties, Glashow, Salam and Weinberg proposed a model based on spontaneously broken non-abelian gauge theories combining electromagnetic and weak interactions. Note that the latter was previously described by the non renormalizable Fermi Lagrangian. This electroweak model was later shown to be renormalizable by 't Hooft and Veltman and the predicted vector bosons were observed at CERN in 1983. Spontaneous symmetry breaking yielding massive vector bosons is achieved through the introduction of the Higgs field, still to be experimentally discovered at the time of writing (mid 2008). 

\item
\underline{\bf Strong interactions}

In the seventies, non-abelian gauge theories with a limited number of fermions were shown to be asymptotically free, meaning that the scale dependent coupling constant is driven to zero at very high energy. This is in agreement with experiments performed at SLAC where quarks inside the hadrons appear to behave as free particles at high energies. Accordingly, the strong force is also described by a renormalizable field theory, Quantum ChromoDynamics (QCD).  However, at low energy the QCD coupling constant increases thus impeding a perturbative treatment.

\item
\underline{\bf Critical phenomena}

Critical phenomena occur in second order phase transitions where the correlation length tends to diverge, so that many degrees of freedom interact. In the seventies, the seminal work of  Wilson led to an understanding of critical phenomena, making use of techniques of renormalizable QFT.    

\item
\underline{\bf Effective field theories}

Later on, the Wilsonian viewpoint on renormalization shed a new light on high energy phenomenology. Indeed, most of the models currently used are thought of as effective field theories, yielding accurate physical predictions within a certain range of energy. This is also the point of view adopted in string theory: Whereas the latter is valid at an energy of the order 10$^{19}$ GeV (Planck's energy) or beyond, the standard model coupled to gravity is understood as a low energy effective action valid at a few hundred GeV. This is just as Fermi's theory that can be understood as a low energy effective theory derived from the electroweak model.

\item
\underline{\bf Constructive QFT}

In fact, the renormalization program in QFT we previously outlined only makes sense at the pertubative level. This means that physical quantities are computed as asymptotic series in a small coupling constant but nothing is said about the  convergence of this expansion. Constructive QFT  is a branch of mathematical physics aiming at defining the theory using general principles and the information encoded in its perturbative expansion (see \cite{constructive}). Here too, the Wilsonian point of view led to substantial progress.  

\item
\underline{\bf Stochastic dynamics}

QFT appear to be a powerful tool each time a system is modeled by a partial differential equation involving a random driving force. For example, the random growth of a surface is described by the Kardar-Parisi-Zhang equation whose solution is expressed as a path integral (see \cite{zee} for a nice pedagogical introduction). Then, renormalization theory provides some efficient tools to investigate the universal behavior of such a system.

\end{itemize}

The ubiquity of renormalization theory  in the previous situations can be understood as follows. In each case, the physical description is supposed to be valid on a wide range of length scales, including extremely small ones that are not directly probed by experiments. Then, renormalization is the set of methods that prescribes how the parameters of the theory have to be adapted to the degrees of freedom that live on a given scale to take into account the effect of the unobserved degrees of freedom that live on smaller scale (or higher energy).

\subsection{Coarse graining for spin systems}

The idea that renormalization is an adaptation of the theory to its natural scale mostly follows from the work of Wilson on critical phemomena. To illustrate how the Wilsonian point of view appeared in the study of critical phenomena in statistical mechanics, let us consider a system of spins ${\sigma_{i}}$  located on the sites of a lattice of spacing $a$. To simplify the discussion much as possible, let us assume that the spins can only take  $q$ different values on each site with a Hamiltonian $H$ involving short range interactions (say a nearest   neighbor one) that favors spin alignment. For instance, one can take
\begin{equation}
H(\sigma)=-J\sum_{\langle i,j\rangle}\delta_{\sigma_{i},\sigma_{j}},
\end{equation}
where the sum runs over all pairs of nearest  neighbors and $J$ is a positive constant. This is the celebrated Ising model for $q=2$ and the $q$-state Potts model in the general case.

The basic object of interest is the partition function
 \begin{equation}
{\cal Z}=\mathop{\sum}\limits_{\sigma}\mathrm{e}^{-\beta H(\sigma)},\label{partition}
\end{equation}
where the sum runs over all spin configurations, $\beta=\frac{1}{kT}$, with $k=1.381\,10^{-23}$ J$\cdot$K$^{-1}$ Boltzmann's constant and $T$ the temperature. For these models,   there is a competition between the tendency of the spin to align in a way to minimize their energy and thermal fluctuations which tend to favor random orientations. Some specific systems like the two dimensional Ising model exhibit a phase transition with a high temperature $T>T_{c}$ disordered phase and a low temperature $T<T_{c}$ ordered one, with $T_{c}$ the critical temperature. As the temperature approaches the critical one, the correlation length tend to diverge, thus leading to long range collective phenomena. Moreover, its has been observed that critical systems fall into universality classes made of systems that have a common asymptotic behavior. For instance, in a given universality class, the correlation length may diverge as $T\rightarrow T_{c}$ with the same exponent,
\begin{equation}
\xi\sim\frac{1}{|T-T_{c}|^{\nu}}.
\end{equation}

\begin{figure}
\begin{center}
\includegraphics[width=12cm]{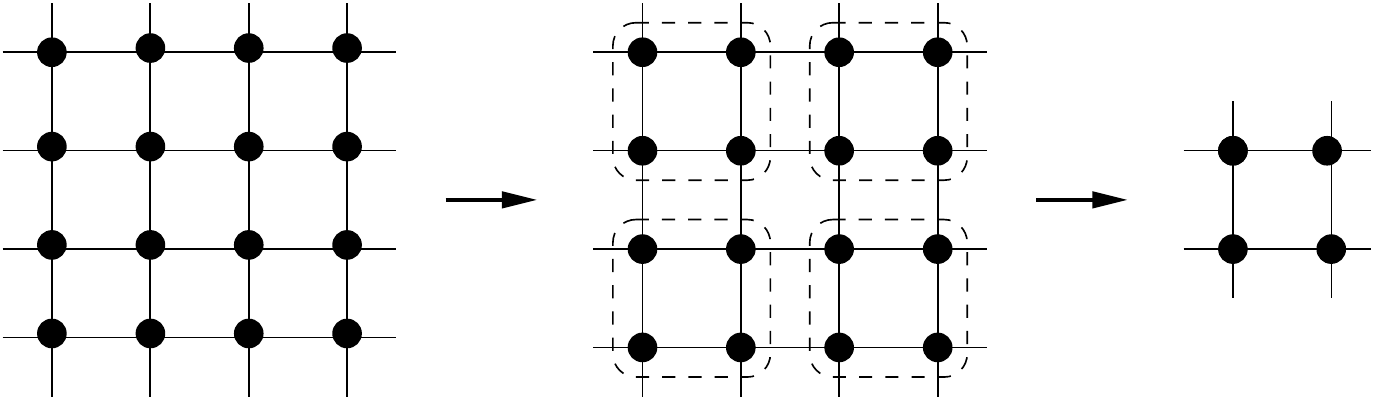}
\caption{Coarse graining}
\label{block}
\end{center}
\end{figure}

Wilson's renormalization group idea  amounts to perform the summation in \eqref{partition} step by step in order to end up with a much simpler description. Specifically, one successively averages over spin configurations  on lattices of increasing spacing (see figure \ref{block})
\begin{equation}
\sigma\xrightarrow{\mathcal{T}}\sigma'
\end{equation}
together with a renormalization group transformation of the Hamiltonian
\begin{equation}
H\xrightarrow{{\cal R}}H'.
\end{equation}
The latter is defined by
\begin{equation}
\mathrm{e}^{-\beta H'(\sigma')}=
\mathop{\sum}\limits_{T(\sigma)=\sigma'}\mathrm{e}^{-\beta H(\sigma)},
\end{equation}
in such a way that the partition function remains invariant
\begin{equation}
{\cal Z}=\mathop{\sum}\limits_{\sigma}\mathrm{e}^{-\beta H(\sigma)}
=\mathop{\sum}\limits_{\sigma'}\mathrm{e}^{-\beta H'(\sigma')}.
\end{equation}

As we perform the coarse graining transformation from  $\sigma$ to $\sigma'$, the lattice size is doubled and it is convenient to also resize the unit of length by a factor of $s=2$, so as to stay on the same lattice. Accordingly, the correlation length is divided by $s$ and after sufficiently many iterations of the renormalization group transformation, one can hope to end up with a tractable problem with a correlation length of the order of the lattice spacing.  Moreover, the sequence of Hamiltonians defined by
\begin{equation}
H_{n+1}={\cal R}(H_{n})
\end{equation}
undergoes some drastic simplifications if it converges towards a fixed point $H^{*}$ of ${\cal R}$. Let us further assume that the linearization of ${\cal R}$ in the vicinity of $H^{*}$ can be diagonalized with eigenvalues that are written in the form $\lambda_{i}=s^{y_{i}}$ and eigenvectors ${\cal O}_{i}$. Starting with a Hamiltonian close to the fixed point, 
\begin{equation}
H=H^{*}+\sum t_{i}{\cal O}_{i},
\end{equation}
the iterations of ${\cal R}$ lead us to distinguish to three cases.

\begin{itemize}
\item
Relevant terms associated with 
$y_i>0$ tend to grow and one has to fine tune the initial Hamiltonian very close to the critical surface (i.e. the basin of attraction of the fixed point) in order to avoid divergences. Relevant terms govern the deviation from criticality and there is usually one of them proportional to $T-T_{c}$.

\item
Irrelevant terms corresponding to $y_{i}<0$ die off very quickly and do not play any significant role.  Information pertaining to them is erased so that universality is recovered.
\item
Marginal terms with zero eigenvalue evolve very slowly but after a deeper study they finally end up in one of the two previous classes, apart from some exceptional cases.
\end{itemize}   

As an elementary illustration, let us determine the critical behavior of the correlation length as an inverse power of $|T-T_{c}|$. We start with a system close to criticality, so that its relevant parameter $t=T-T_{c}$ with eigenvalue $y$ is very small and its correlation length $\xi$ large. We now iterate $n$ times the renormalization group transformation in such a way that the relevant parameter $t'$ becomes of order unity and the correlation length $\xi'$ close to the lattice spacing,
\begin{equation}
a\sim\xi'\sim\frac{\xi}{s^{n}}\quad\mathrm{and}\quad1\sim t'\sim s^{ny}t.
\end{equation}
Eliminating $s^{n}$ we obtain the critical behavior of the correlation, 
\begin{equation}
\xi\sim\frac{1}{|T-T_{c}|^{\frac{1}{y}}},
\end{equation}
where we have discarded a slowly varying  multiplicative constant .

To end this very brief introduction to Wilson's ideas, let us sketch a renormalization group proof of the central limit theorem in probability theory, which has the merit of explaining the universal nature of the Gau\ss ian probability law (see \cite{sinai} and \cite{jona}). Consider $N=2^{n}$ independent identically distributed (IID) random variables $\xi_{i}$ with probability density $\rho_{0}$, assuming that its variance is 1 and a vanishing expectation value. Recall that the central limit theorem states that  
\begin{equation}
\zeta_{N}=\frac{\xi_{1}+\cdots+\xi_{N}}{\sqrt{N}}
\end{equation}
converges in law towards the normal law with density $\frac{1}{\sqrt{2\pi}}\mathrm{e^{-\frac{x^{2}}{2}}}$ as $N\rightarrow\infty$.

In the spirit of coarse graining, let us compute $\zeta_{N}$ by iterating $n$ times the transformation that associates to two IID random variables $\xi'$ and $\xi''$ the new one defined by  
\begin{equation}
\xi=\frac{\xi'+\xi''}{\sqrt{2}}.
\end{equation}
The probability density ${\cal R}(\rho)$ of $\xi$ is expressed in terms of that of $\xi'$ and $\xi''$ as
\begin{equation}
\left[{\cal R}(\rho)\right](x)=\sqrt{2}\int dy\, \rho(y)\rho(\sqrt{2}x-y),
\end{equation}
so that the probability density of $\zeta_{N}$ can be written as
\begin{equation} 
\rho_{n}={\cal R}^{n}(\rho_{0}).
\end{equation}
The renormalization map ${\cal R}$ admits the Gau\ss ian
\begin{equation}
\rho_{*}(x)=\textstyle{\frac{1}{\sqrt{2\pi}}}\mathrm{e}^{-x^{2}/2},
\end{equation}
as a fixed point. To investigate the stability of this fixed point, let us linearize ${\cal R}$ in the neighborhood of the Gau\ss ian. Writing the perturbation as
\begin{equation}
\rho(x)=\textstyle{\frac{1}{\sqrt{2\pi}}}\mathrm{e}^{-x^{2}/2}
\Big(1+\epsilon(x)\Big),
\end{equation}
we have, up to order $\epsilon^{2}$ terms, 
\begin{equation}
{\cal R}[\rho](x)={\textstyle\frac{1}{\sqrt{2\pi}}}\mathrm{e}^{-x^{2}/2}\left[1+{\textstyle\frac{2}{\sqrt{\pi}}}
\int dy \,\mathrm{e}^{-y^{2}}\epsilon\left(y+x/\sqrt{2}\right)
\right]
+O(\epsilon^{2}).
\end{equation}
The eigenvectors of the linearization  of ${\cal R}$ are obtained by taking $\epsilon$ proportional to a Hermite polyomial $H_{n}$. This is easy to check using their generating function
\begin{equation}
\mathrm{e}^{-\frac{s^{2}}{2}+xs}=\sum_{n}H_{n}(x)\,\frac{s^{n}}{n!}.
\end{equation}
The associated eigenvalues are $\lambda_{n}=2^{1-\frac{n}{2}}$, so that all the terms involving Hermite polynomials of order $n>2$ are irrelevant. For $n=0,1,2$, the coefficients of the expansion of $\epsilon$  have to vanish in order to make sure that the iteration of ${\cal R}$ takes the probability density towards the fixed point. They simply state that the probability distribution has to be normalized ($n=0$), with vanishing expectation value ($n=1$) and with variance  1 ($n=2$). In fact, only $n=0$ and $n=1$ is are relevant, $n=2$ is marginal. Imposing these conditions on the probability density we start with is analogous to the fixing the otherwise divergent quantities in QFT by low energy measurements.

\subsection{Wilsonian renormalization in QFT }

The very same ideas can be applied to QFT, with the partition function replaced by the path integral
\begin{equation}
{\cal Z}=\int[D\phi]_{\Lambda}\,\mathrm{e}^{-S[\phi]}.
\end{equation}   
The integration  measure involves an UV cut-off $\Lambda$, restricting the integration over fields whose Fourier modes vanish for momenta above $\Lambda$. It is analogous to a system defined on a lattice and reflects the fact that rapid variations of the fields on spatial scales below $a\sim\frac{1}{\Lambda}$ do not make sense in this context.  

In analogy with spin blocking, one integrates over fast modes in order to obtain a equivalent field theory with a lower cut-off $\Lambda'=\frac{\Lambda}{s}$. To proceed, let us separate the field $\phi$ into its fast component $\phi''$ with momenta between $\Lambda'$ and $\Lambda$ and its slow one $\phi'$ with momenta below $\Lambda'$. By performing the integration over $\phi''$, we obtain the low energy effective action 
\begin{equation}
\mathrm{e}^{-S'[\phi']}=\int[D\phi'']_{\Lambda',\Lambda}\,
\mathrm{e}^{-S[\phi'+\phi'']},
\end{equation}
The partition function may be computed either with $S$ or with $S'$,
\begin{equation}
{\cal Z}=\int[D\phi]_{\Lambda}\,\mathrm{e}^{-S[\phi]}=
\int[D\phi']_{\Lambda'}\,\mathrm{e}^{-{\cal S'}[\phi']}.
\end{equation}
The equality between the two terms follows form a substitution of $S'$ in terms of $S$ followed by the change of variables $\phi=\phi'+\phi''$. The remaining integration is cancelled by a normalization factor. These are self-understood constants in all formulae that do not depend on the actions and are defined by imposing that effective actions and partition functions are trivial for trivial actions. 

Besides, this remains true if we insert in the path integral any functional of $\phi'$, so that one can conclude that low energy  physics remains unchanged if we simultaneously lower the cut-off and replace $S$ by ${\cal S'}$. All the low energy effects of the physics at the intermediate scale  $\Lambda'$ have been encoded in $S'$, which may therefore have a rather complicated form even if $S$ is simple. 

\begin{figure}
\begin{center}
\includegraphics[width=10cm]{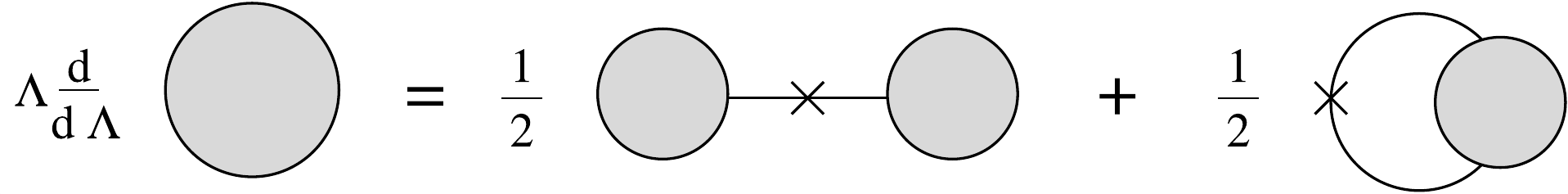}
\caption{Diagrammatic form of Polchinski's equation}
\label{Polchinskidiag}
\end{center}
\end{figure}

As for spin systems, one is led to iterate the renormalization group transform to obtain a sequence of effective actions with decreasing cut-offs. In the context of QFT, it more transparent to adopt a continuous formulation, leading to a family of cut-off dependent effective actions $S_{\Lambda}$, governed by a differential equation     
\begin{equation}
\Lambda\frac{dS_{\Lambda}}{d\Lambda}=\beta(\Lambda,S_{\Lambda}).
\label{RG}
\end{equation}
The action with started with is encoded in the initial condition, $S_{\Lambda_{0}}=S_{0}$, valid at a very high energy scale $\Lambda_{0}$. To derive the renormalization group equation \eqref{RG}, let us note that $S_{\Lambda}$ must be obtained from $S_{\Lambda_{0}}$ by integrating over all the modes between $\Lambda$ and $\Lambda_{0}$. This is achieved in the path integral approach by
\begin{equation}  
\mbox{e}^{-S_{\Lambda}[\phi]}= \int\, [D\chi]\,\mathrm{e}^{-{\textstyle\frac{1}{2}}\chi\cdot A^{-1}_{\Lambda,\Lambda_{0}}\cdot\chi-S_{\Lambda_0}[\chi+\phi]},
\end{equation}
where $A_{\Lambda,\Lambda_{0}}$ is a smooth cut-off dependent propagator that implements a smooth analogue of the integration over the fields with Fourier modes between $\Lambda$ and $\Lambda_{0}$. For example, we shall often use a cut-off inspired by Schwinger's representation
\begin{equation}
{A_{\Lambda,\Lambda_{0}}}(p,q)=\delta(p,q)\int^{\frac{1}{\Lambda^{2}}}_{\frac{1}{(\Lambda_{0})^{2}}}\!\!d\alpha\,\,
\mathrm{e}^{-\alpha p^{2}},
\end{equation}
with the mass term relegated in the effective action. Note that $\Lambda$ plays the role of an IR cut-off whereas $\Lambda_{0}$ is an UV one. By differentiating with respect to $\Lambda$, we get Polchinski's equation (see \cite{Polchinski}), 
\begin{equation}
\Lambda\frac{\partial
S_{\Lambda}}{\partial\Lambda}=\frac{1}{2}\int\!dp dq\,
\Lambda\frac{\partial
A_{\Lambda,\Lambda_{0}}}{\partial\Lambda}(p,q)\left(
\frac{\delta^{2}S}{\delta\widetilde{\phi}(p)\delta\widetilde{\phi}(q)}- \frac{\delta
S}{\delta\widetilde{\phi}(p)} \frac{\delta S}{\delta\widetilde{\phi}(q)} \right),
\label{erge}
\end{equation}
which is conveniently captured by Feynman diagrams as illustrated on figure \ref{Polchinskidiag}. Although $A_{\Lambda,\Lambda_{0}}$ depends on both the UV and IR cut-offs,  its derivative only depends on $\Lambda$. 

Here we have use the Fourier transform $\widetilde{\phi}(p)$ of the field $\phi(x)$ but an analogous equation can be written in position space,
\begin{equation}
\Lambda\frac{\partial
S_{\Lambda}}{\partial\Lambda}=\frac{1}{2}\int\!dxdy\,
\Lambda\frac{\partial
A_{\Lambda,\Lambda_{0}}}{\partial\Lambda}(x,y)
\left(
\frac{\delta^{2}S}{\delta\phi(x)\delta\phi(y)}- 
\frac{\delta S}{\phi(x)}\frac{\delta S}{\delta\phi(y)}\right),
\end{equation}
with 
\begin{equation}
A_{\Lambda,\Lambda_{}0}(x,y)=\int_{{\Bbb R}^{D}}\frac{d^{D}p}{(2\pi)^{D}}\,
\mathrm{e}^{\mathrm{i}p\cdot(x-y)}\,A_{\Lambda,\Lambda_{0}}(p,-p).
\end{equation}

To close this terse section on renormalization, let us draw a few high energy physics consequences from the Wilsonian approach. In this context, $\Lambda_{0}$ is the UV cut-off introduced in order to render the Feynman diagrams convergent and $S_{0}$ identifies with the bare action. $S_{\Lambda}$ is a low energy effective action, which has to be further used in Feynman diagrams having all their internal momenta below a fixed low energy scale $\Lambda$. As we try to get rid of $\Lambda_{0}$ by taking it to infinity, some of the parameters entering in $S_{\Lambda}$ tend to diverge. The latter have to be fixed by low energy measurement and renormalization simply amounts to choosing a bare action $S_{0}$ in such a way that the flow will meet these low energy conditions. Since the limit $\frac{\Lambda_{0}}{\Lambda}\rightarrow\infty$ corresponds to an iteration of a large number of renormalization group transforms, this is very close to what has been done for critical phenomena. From this point of view, one can understand the ubiquity of renormalizable theories depending on a few paramaters as an analogue of universality in statistical mechanics. Whatever $S_{0}$ we start with, provided it is natural (i.e. its parameters are numbers $\simeq 1$ times powers of $\Lambda_{0}$ dictated by dimensional analysis), we always end up with up with renormalizable theories at energies $\Lambda\ll \Lambda_{0}$.  By the same token, universality implies that the renormalized theory is independent of way we implement the cut-off.

Besides, this analogy can be pushed further: All the renomalizable interactions that appear in high energy physics can be thought of as low energy effective theories derived using  a renormalization group flow from a yet unknown theory, valid at a very high energy scale $\Lambda_{0}$. The latter theory therefore plays in particle physics a role analogue to lattice models in condensed matter physics. As such, it may not even be a field theory and rely on some utterly new ideas, like string theory, spinfoam models or noncommutative geometry. The important point is that it comes equipped with a high energy cutoff $\Lambda_{0}$, of the order of Planck's length, which is physical and should not be taken to infinity. At energies slightly  below $\Lambda_{0}$, the physics is well captured by a QFT whose field content, action and symmetries are postulated, but should determined by the unknown theory. Starting with this action, we move towards  lower energies $\Lambda$ using the renormalization group equation and end up with an effective action determined by a few relevant parameters to be experimentally determined. Irrelevant terms corresponding to non renormalizable Lagrangians are also present, though very small, and are innocuous because the cut-off remains finite. Therefore, there is no special problem with gravity, the latter being simply the leading irrelevant correction to the Standard Model, and the Einstein-Hilbert Lagrangian  may be itself corrected by higher order terms. Of course this does by no means solve the problem of quantum gravity, the later presumably amounts to finding the unknown theory we should start with. 

One of the main lessons from the Wilsonian approach to renormalization in high energy physics is that the cut-off acquires a physical meaning: It sets the upper limit of validity of the theory. This is in sharp contrast with the point of view adopted a few decades ago where the cut-off was a mere scaffolding needed to build the theory and that has to be removed as soon as possible.

This approach also raises another question: How do we make sure that the relevant paramaters, like the masses, are small and not of the order of $\Lambda_{0}$? For lattice systems, an operator can tune the bare Hamiltonian sufficiently close to the critical surface so that we end up with finite results even after a large number of iterations. In particle physics, finding such a process remains an open question and is  particularly important for a scalar boson like the Higgs field. It seems that such a problem can be solved by introducing supersymmetric particles but at the tome of writing (mid 2008) no such particles have been found.

\section{Rooted trees and power series of non linear operators}

\subsection{Rooted trees in numerical analysis}

Soon after Connes and Kreimer presented an algebraic framework base on rooted trees for renormalization \cite{ck0}, it was realized by Brouder \cite{brouder} that perturbative renormalization  has some intriguing connection with the algebraic techniques used in the numerical analysis of differential equations.

\begin{figure}
\begin{center}
\includegraphics[width=2cm]{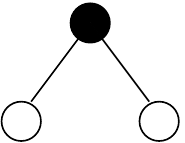}
\caption{A rooted tree with one branching}
\label{treeexample}
\end{center}
\end{figure}

To begin with, let us expand in powers of $s-s_{0}$ the solution of the differential equation in ${\Bbb R}^{n}$, 
\begin{equation}
\frac{dx}{ds}=X(x),\qquad x(s_{0})=x_{0},\label{diffeq}
\end{equation}
as
\begin{equation}
x(s)=\sum_{n}\frac{(s-s_{o})^{n}}{n!}\frac{d^{n}x}{ds^{n}}\Big|_{s=s_{0}}.
\end{equation}
To proceed, we compute the successive derivatives of $x$ with respect to $s$,
\begin{eqnarray}
\frac{dx^{i}}{ds}&=&X^{i}\cr
\frac{d^{2}x^{i}}{ds^{2}}&=&\sum_{j}\frac{\partial X^{i}}{\partial x^{j}}\,X^{j}\cr
\frac{d^{3}x^{i}}{ds^{3}}&=&\sum_{j,k}\frac{\partial X^{i}}{\partial x^{j}}
\,\frac{\partial X^{j}}{\partial x^{k}}\,X^{k}
+\frac{\partial^{2}X^{i}}{\partial x^{j}\partial x^{k}}\,X^{j}X^{k}\cr
\frac{d^{4}x^{i}}{ds^{4}}&=&\sum_{j,k,l}\frac{\partial X^{i}}{\partial x^{j}}\,
\frac{\partial X^{j}}{\partial x^{k}}\,\frac{\partial X^{k}}{\partial x^{l}}\,X^{l}
+3\,\frac{\partial^{2}X^{i}}{\partial x^{j}\partial x^{k}}
\frac{\partial X^{k}}{\partial x^{l}}\,\,X^{j}X^{l}\cr
&&\quad+\frac{\partial^{3} X^{i}}{\partial x^{j}\partial x^{k}\partial x^{l}}
\,X^{j}X^{k}X^{l}
+\frac{\partial X^{i}}{\partial x^{j}}\frac{\partial^{2}X^{j}}{\partial x^{k}\partial x^{l}}\,X^{k}X^{l}
\label{expand}
\end{eqnarray}
All the terms in this expansion turn out to be in one-to-one correspondence with rooted trees with at most four vertices. For example, the second term in the derivative $\frac{d^{2}x}{dt^{2}}$ corresponds to the tree depicted in figure \ref{treeexample}.

This is familiar from Runge-Kutta methods in numerical analysis. Indeed, a Runge-Kutta method is defined by a square matrix $(a_{ij})_{1\leq i,j\leq n}$ and a vector $(b_{i})_{1\leq i\leq n}$ of real numbers, arranged into a table
\begin{equation}
\begin{array}{c|cccc} 
c_{1}&a_{11}&a_{12}&\cdots&a_{1n}\cr
c_{2}&a_{21}&a_{22}&\cdots&a_{2n}\cr
\vdots&\vdots&\vdots&&\vdots\cr
c_{n}&a_{n1}&a_{n2}&\cdots&a_{nn}\cr
\hline
&b_{1}&b_{2}&\cdots&b_{n}
\end{array}
\label{RK}
\end{equation}
where $c_{i}=\sum_{j}a_{ij}$.  Then, the solution of the differential equation
\begin{equation}
\frac{dx}{ds}=X(x),\quad\mathrm{with}\quad x(s_{0})=x_{0},\label{diffeqRK}
\end{equation}
is given at $s_{1}=s_{0}+h$ by
$
x_{1}=x_{0}+h\sum_{i=1}^{n}b_{i}X(y_{i}),
$
where $y_{i}$ is determined by
$
y_{i}=x_{0}+h\sum_{j=1}^{n}a_{ij}X(y_{j}).
$
When $a_{ij}=0$ for $j>i$, the computation of $y_{i}$ in terms of trees is straightfoward. Then, trees turn out to be useful to compare the result of Runge-Kutta method with the expansion  \eqref{expand}.

It is an amazing fact discovered by Butcher that Runge-Kutta methods can be composed \cite{butcher}. One can perform a first approximate computation using a first method to obtain $x$ as a function of the initial condition $x_{0}$.   Then, we can consider $x$ as a new initial condition and proceed to a different approximate computation resulting in $x'$. This amounts to a single computation of $x'$ with initial condition $x_{0}$, using a third method which is the product of the previous ones.

\subsection{Hopf algebra structure}

The occurrence of rooted trees in numerical analysis through Runge-Kutta methods and their composition law is conveniently described using the Hopf algebra of rooted trees. To begin with, recall that a tree is a connected diagram without loop\footnote{We use the QFT terminology: A loop is what graph theorists call a cycle.}  and a rooted tree is a tree with a distinguished vertex called the root. A rooted tree is oriented from the root to its terminal vertices, the leaves. Except in the very last paragraph devoted to a bijection between planar diagrams and trees, we always consider isomorphism classes of trees.

Let ${\cal H}_{T}$ be the commutative algebra generated by all (isomorphism classes of) rooted trees. It admits a Hopf algebra structure defined on the generators as follows.

\begin{itemize}
\item
The coproduct $\Delta:\,{\cal H}_{T}\rightarrow{\cal H}_{T}\otimes{\cal H}_{T}$ is
\begin{equation}
 \Delta(t)=t\otimes 1+1\otimes t+
\mathop{\sum}\limits_{c\;\mathrm{admissible}\;\mathrm{cut}}
P_{c}(t)\otimes R_{c}(t)
\end{equation}
A cut of an edge is admissible cut if any path from any leaf to the root is cut at most once. $R_{c}(t)$ is the connected component that contain the root after the cut, it is made of vertices located above the cut. $P_{c}(t)$ is the product of remaining trees corresponding to vertices below the cut with as new root those vertices immediately below the cut. For example, for the simplest tree  with one branching, we have

%\begin{figure*}[h!]
%\centerline{\hspace{0.5cm} 
%\(
\begin{equation}
\Delta\left(
\parbox{0.8cm}{\mbox{\includegraphics[width=0.8cm]{t3.pdf}}}\right)=1\otimes
\parbox{0.8cm}{\mbox{\includegraphics[width=0.8cm]{t3.pdf}}}+
\parbox{0.8cm}{\mbox{\includegraphics[width=0.8cm]{t3.pdf}}}\otimes 1 + 2
\parbox{0.2cm}{\mbox{\includegraphics[width=0.2cm]{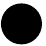}}} \otimes
\parbox{0.2cm}{\mbox{\includegraphics[width=0.2cm]{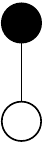}}}+
\parbox{0.2cm}{\mbox{\includegraphics[width=0.2cm]{t1.pdf}}}
\parbox{0.2cm}{\mbox{\includegraphics[width=0.2cm]{t1.pdf}}}\otimes
\parbox{0.2cm}{\mbox{\includegraphics[width=0.2cm]{t1.pdf}}}. 
\label{examplecop}
\end{equation}
%\)}
%\end{figure*}

\item The counit $\epsilon:\,{\cal H}_{T}\rightarrow{\Bbb C}$ is trivial, $\epsilon(t)=0$ unless $t=1$.

\item
The antipode $S:\,{\cal H}_{T}\rightarrow{\cal H}_{T}$ is defined as a sum over all cuts,
\begin{equation}
S(t)= -\mathop{\sum}\limits_{c\;\mathrm{cut}}
(-1)^{n_{c}(t)}\Pi_{c}(t), \label{antipodetree}
\end{equation}
with $n_{c}(t)$ the number of edges of the cut.
\end{itemize}
Since the Hopf algebra ${\cal H}_{T}$ is commutative, its characters form a  group $G_{T}$ for
the convolution product
\begin{equation}
\alpha\ast\beta=(\alpha\otimes\beta)\circ\Delta
\end{equation}
with unit $\epsilon$ and inverse $\alpha^{-1}=\alpha\circ S$. Recall that a character is an algebra morphism from ${\cal H}_{T}$ into a fixed commutative ring,
\begin{equation}
\left\{
\begin{array}{rcl}
\alpha(\lambda a+\mu b)&=&\lambda\,\alpha(a)+\mu\,\alpha(b)\cr
\alpha(ab)&=&\alpha(a)\alpha(b),
\end{array}
\right.
\end{equation}
for any $a,b\in{\cal H}_{T}$ and $\lambda,\mu\in{\Bbb C}$.   Let us emphasize that the characters must take their values in a commutative ring, in order to check that that the product of characters is still a character. Most of the time we work with complex numbers but it may be useful to consider rings of functions.

\begin{figure}
\begin{center}
\includegraphics[width=2cm]{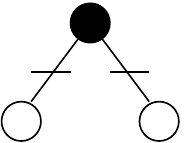}
\caption{An admissible cut on the rooted tree of figure \ref{treeexample}}
\label{treeexamplecut}
\end{center}
\end{figure}

A character of ${\cal H}_{T}$ is entirely specified by its values on the trees, which may be arbitrary, so that it encodes an infinite sequence of complex numbers. In fact, ${G}_{T}$ is a Lie group whose Lie algebra ${\cal G}_{T}$ consists in all infinitesimal characters. The latter are linear maps $\delta:\,{\cal H}_{T}\rightarrow{\Bbb C}$, satisfying the infinitesimal form of multiplicativity, 
\begin{equation}
\delta(ab)=\epsilon(a)\delta(b)+\delta(a)\epsilon(b)
\end{equation}
for any $a,b\in{\cal H}_{T}$. Alternatively, infinitesimal characters can be defined as derivations with values in the trivial bimodule provided by $\epsilon$.

The convolution exponential $\exp_{\ast}:\,{\cal G}_{T}\rightarrow G_{T}$, defined by
\begin{equation}
\exp_{\ast}(\delta)=\epsilon+\delta+\frac{\delta\ast\delta}{2}+\cdots+
\frac{\delta^{\ast n}}{n!}+\cdots,
\end{equation}
and its inverse $\log_{\ast}{G}_{T}\rightarrow {\cal G}_{T}$, given by
\begin{equation}
\log_{\ast}(\alpha)=
\alpha\!-\!\epsilon-\frac{(\alpha\!-\!\epsilon)\ast(\alpha\!-\!\epsilon)}{2}+\cdots+
(-1)^{n-1}\frac{(\alpha\!-\!\epsilon)^{\ast n}}{n}+\cdots,
\end{equation}
allow to identify ${\cal G}_{T}$ as the infinitesimal form of $G_{T}$. Convergence of these series is not an issue, because both $\delta$ and $\alpha-\epsilon$ vanish on 1, 
so that only a finite number of terms survive when evaluated on a tree $t$. Because infinitesimal characters vanish on products of trees,  they often only involve very simple combinatorial problems.  Then, less trivial identities can be obtained by exponentiation. We shall illustrate this fact on several examples in the sequel.

To investigate the the structure of the group of characters $G_{T}$, it is convenient to first define a decreasing sequence $K_{T}^{n}$ of subgroups of $G_{T}$ by requiring that elements of $K_{T}^{n}$ are characters that vanish on all trees of order $\leq n$. This is a sequence of normal subgroups of $G_{T}$, so that one can define an increasing sequence of quotient groups $G^{n}_{T}=G_{T}/K_{T}^{n+1}$.  $G_{T}^{n}$ is a finite dimensional nilpotent Lie group with Lie algebra ${\cal G}_{T}^{n}$ and one recovers $G_{T}$ and its Lie algebra as the projective limits of the corresponding sequences.  Besides, one can construct recursively these groups since $G_{T}^{n+1}$ is an abelian extension of $G_{T}^{n}$. Recall that given a group $G$ and an abelian group $K$, a abelian extension of $G$ by $K$ is a new group $\tilde{G}$ defined as follows. As a set $\tilde{G}=G\times K$, but its product law differs from the direct product law by a 2-cocycle $\theta$,
\begin{equation}
\left(\alpha,\lambda\right)\cdot\left(\beta,\mu\right)=
\left(\alpha\beta,\lambda+\mu+\theta(\alpha,\beta)\right),
\end{equation}
for any $\alpha,\beta\in G$ and $\lambda,\mu\in K$. The associativity of the multiplication on $\tilde{G}$, is ensured by the cocycle condition
\begin{equation}
\theta(\beta,\gamma)-\theta(\alpha\beta,\gamma)+\theta(\alpha,\beta\gamma)-\theta(\alpha,\beta)=0
\end{equation}
for all $\alpha,\beta\gamma\in G$.

For any rooted tree $t$ of order $n+1$, let us define
\begin{equation}
\theta_{t}(\alpha,\beta)=\sum_{c\,\mathrm{admissible \,cuts\, of\,}t\atop c\neq\mathrm{empty}\;c\neq\mathrm{full}
}\alpha(t_{1})\cdots\alpha(t_{n})
\,\beta\left(\frac{t}{t_{1}\cdots t_{n}}\right),
\end{equation}
where we exclude the empty and the full cuts. Using the reduced coproduct $\Delta'=\Delta-1\otimes\mathrm{Id}-\mathrm{Id}\otimes 1$, the cocycle reads, 
\begin{equation}
\theta_{t}(\alpha,\beta)=\left(\alpha\otimes\beta\right)\circ\Delta' (t),
\end{equation}
and thus identifies with $\Delta'$ considered as a multiplicative function from the algebra generated by trees of order $n+1$ to the tensor product of to copies of the algebra generated by trees of order less or equal to $n$. This allows for a quick check of the cocycle condition since it follows directly from the associativity of $\Delta$. 

${\cal H}_{T}$ is graded by the number of vertices $|t|$ and fits into the general framework of graded and commutative Hopf algebras. Recall that a graded and commutative Hopf algebra is a direct sum ${\cal H}=\oplus_{n\in{\Bbb N}}{\cal H}_{n}$ with
 $
{\cal H}_{m}\cdot {\cal H}_{n}\subset{\cal H}_{mn}
$
and
$
\Delta {\cal H}_{n}\subset\oplus_{m\in{\Bbb N}}{\cal H}_{m}\otimes{\cal H}_{n-m}.
$
This framework allows us to implement inductive reasoning.  If we further assume that ${\cal H}_{0}$ is one dimensional, a general structure theorem \cite{cartier} states that ${\cal H}$ is a free commutative algebra so that all the previous results still hold, since most of the proofs rely on inductive reasoning.

\subsection{Power series of non linear operators}

When solving the differential equation \eqref{diffeq} using a sum indexed by rooted trees, we associate each tree with a combination of the derivative of $X$. In order to generalize this construction, it is fruitful to view trces as indices for formal power series of non linear operators.

To proceed, let us consider any smooth map $X$ from a Banach space ${\cal E}$ to itself and let us define $X$ raised to the power of the tree $t$ as
\begin{equation}
X^{t}=\prod_{v\in t}^{\longrightarrow}X^{(n_{v})}
\end{equation}
where $n_{v}$ is number of edges leaving the vertex $v$ and  $X^{(n)}$ is the  $n^{\mbox{\tiny th}}$ order differential of $X$.  By convention, the tree is oriented from the root to the leaves and we compose the operators following this order. When evaluated at a point $x\in{\cal E}$, $X^{(n)}(x)$ is  $n$-linear map and the product refers to the composition of those linear maps from the root to the leaves. As a result, $X^{t}$ is a smooth map from ${\cal E}$ to itself.
For instance, we have
\begin{equation}
X^{\parbox{0.1cm}{\mbox{\includegraphics[width=0.1cm]{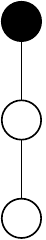}}}}=X'\big[X'[X]\big]\quad\mathrm{and}\quad
X^{\parbox{0.4cm}{\mbox{\includegraphics[width=0.4cm]{t3.pdf}}}}=X''\big[X, X\big].
\end{equation}
Note that this  corresponds exactly to the combinations of partial derivatives used for solving the 
differential equation \eqref{diffeq}. Let us note that there is an interesting similarity with the parenthesized words introduced by Kreimer in the very first paper on the Hopf algebras of renormalization  \cite{kreimer}.

Given an arbitrary character $\alpha\in G_{T}$, let us define the formal power series  
\begin{equation}
\Psi_{\alpha}(X)=\sum_{t}\alpha(t)\,\frac{X^{t}}{\mathrm{S}_{t}},
\end{equation}
with $\mathrm{S}_{t}$ the symmetry factor of the tree, which is the cardinal of its automorphism group. This has to be understood as a formal power series of smooth maps, i.e. when evaluated on any $x\in{\cal E}$ it yields a formal power series of elements of ${\cal E}$, which is ordered in increasing number of vertices of the trees.  Let us also note that the series always starts with the empty tree which is nothing but the unit of the algebra ${\cal H}_{T}$. By convention, any operator raised to the power of the empty tree is the identity, $X^{\emptyset}=\mathrm{id}$.

If we now consider another $\beta\in G_{T}$, then we can compose $\Psi_{\alpha}(X)$ with $\psi_{\beta}(X)$, Taylor expand and reorganize the series in terms of rooted  trees. It turns out that the result is naturally expressed using the product law in $G_{T}$,
\begin{equation}
\underbrace{\Psi_{\alpha}(X)\circ\Psi_{\beta}(X)}_{\mathrm{composition}}=
\underbrace{\Psi_{\beta\,\ast\,\alpha}(X)}_{\mathrm{convolution}}.
\label{composition}
\end{equation}
This composition law is identical to the composiion law of B-series \cite{wanner},
underlying Butcher's work on the algebraic structure of Runge-Kutta methods. To prove this relation, we write $\delta=\log_{\ast}(\alpha)$ and consider both sides of as a function of $s\in{\Bbb R}$ for $\alpha_{s}=\exp_{\ast}(s\delta)$. Then, we derive both sides with respect to $s$ and check that the derivatives agree using simple combinatorial identities. Let us note  that if $X$ is linear, all derivatives $X^{(n)}$ vanish for $n>1$ so that only trees without branchings appear. In this case eq. \eqref{composition} amounts to the ordinary multiplication of power series of a linear operator.

It is also convenient to allow the trees to carry some decorations on their vertices. Then, all the structure remains similar, with $R_{c}(t)$ and $P_{c}(t)$ carrying decorations inherited from $t$ in the definition of the coproduct. The main interest of decorated rooted trees lies in the fact that they allow us to define power series of several non linear operators, each operator being labeled by its decoration.

The most useful series is the geometric one, defined by summing over all trees with a weight one, 
\begin{equation}
(\mathrm{id}-X)^{-1}=\sum_{t}\frac{X^{t}}{\mathrm{S}_{t}}. \label{geometrictree}
\end{equation}
As suggested by the notation, it yields an inverse of $\mathrm{id}-X$ for the composition law. Accordingly, it provides us with a perturbative solution of the fixed point equation
\begin{equation}
x=x_{0}+X(x)\label{fixedpoint}
\end{equation}
as a sum over trees
\begin{equation}
 x=(\mathrm{id}-X)^{-1}(x_{0})=\sum_{t}\frac{X^{t}}{\mathrm{S}_{t}}(x_{0}).
\end{equation}
Moreover, let us note that the tree like structure of the geometric expansion proves to be extremely helpful in dealing with convergence issues. To proceed, let us assume that the $n^{\mathrm{th}}$ order differential is  a bounded $n$-linear map such that $||X^{(n)}(x_{0})||\leq a_{n}$, with $a_{n}$ a sequence of real numbers such that
\begin{equation}
f(z)=\sum_{n=0}^{\infty}\frac{a_{n}}{n!}\,z^{n}
\end{equation}
converges. Then, the series 
\begin{equation}
 x=(\mathrm{id}-hX)^{-1}(x_{0})=\sum_{t}h^{|t|}\frac{X^{t}}{\mathrm{S}_{t}}(x_{0}).
\end{equation}
is bounded in norm by the series associated to the equation $z=hf(z)$, which is known to be an analytic function of $h$ at the origin by the implicit function theorem. 

Beyond the geometric series, it may also be useful to consider the binomial series.  For any character $\alpha$ and complex number $a$, let us define
\begin{equation}
(\alpha)^{\ast a}=\epsilon+
a(\alpha\!-\!\epsilon)+\frac{a(a-1)}{2}(\alpha\!-\!\epsilon)\ast(\alpha\!-\!\epsilon)+\cdots+
\frac{a(a-1)\cdots(a-n+1)}{n!}(\alpha\!-\!\epsilon)^{\ast n}+\cdots.\label{binomial}
\end{equation}
Obviously, the binomial series fulffils all the identities we expect
$(\alpha)^{\ast a}\ast(\alpha)^{\ast b}=(\alpha)^{\ast a+b}$ and $((\alpha)^{\ast a})^{\ast b}=(\alpha)^{\ast ab}$ at the notable exception of $(\alpha)^{\ast a}\ast (\beta)^{\ast a}=(\alpha\ast\beta)^{\ast a}$, which does not hold since  the convolution product is not commutative.  

If we take $\alpha$ to be the character that takes the value $1$ on the tree with one vertex and vanishes on any other non trivial tree, then the iterations of the coproduct are easy to compute,
\begin{equation}
(\alpha)^{\ast a}(t)=\sum_{n=d_{t}}^{|t|}N(n,t)\frac{a(a-1)\cdots (a-n+1)}{n!},
\end{equation}
where $N(n,t)$ is the number of surjective maps form the vertices of $t$ to $\left\{1,\cdots,n\right\}$, strictly increasing from the root to the leaves. $d_{t}$ is the depth of the tree which is the length of the longest path from the root to the leaves. 

Accordingly, one can apply the binomial series to a non linear operator,
\begin{equation}
(\mathrm{id}+X)^{a}=\sum_{t}
\sum_{n=d_{t}}^{|t|}N(n,t)\frac{a(a-1)\cdots (a-n+1)}{n!}
\frac{X^{t}}{\mathrm{S}_{t}}\label{binomX}
\end{equation}
the coefficients in from of $\frac{X^{t}}{\mathrm{S}_{t}}$ can be considered as tree-lie generalizations of the binomial coefficients, to which they reduce on trees without branchings.
For instance, up to order 4 we have
\begin{equation}
(\mathrm{id}+X)^{a}=1+aX+\frac{a(a-1)}{2}\left(X'[X]+\frac{1}{2}X''[X,X]+\right)+\frac{a(a-1)(a-2)}{6}\bigg( X'[X'[X]]+X''[X,X]\bigg)+\dots.
\end{equation}
This allows, for instance, to compute the square root of a diffeomorphism 
\begin{eqnarray}
\renewcommand{\baselinestretch}{4}
\sqrt{\mathrm{id}+X}&=&\frac{1}{2}X-\frac{1}{8}X'[X]+\frac{1}{16}X'[X'[X]]-\frac{5}{128}X'[X'[X'[X]]]\cr
&&+\frac{1}{128}X''[X,X'[X]]-\frac{1}{2\cdot 64}X'[X''[X,X]]+\frac{1}{6\cdot 64}X'''[X,X,X]+\cdots
\end{eqnarray}
which fulfills $\sqrt{\mathrm{id}+X}\circ\sqrt{\mathrm{id}+X}=\mathrm{id}+X$ up to terms of fifth order in $X$. 

Let us also introduce   the number $\widetilde{N}(n,t)$ of surjective maps form the vertices of $t$ to $\left\{1,\cdots,n\right\}$, increasing (but not necessarily in a strict manner) from the root to the leaves.
Then, by applying the binomial series to the geometric one, we get
\begin{equation}
(\mathrm{id}-X)^{-a}=\left((\mathrm{id}-X)^{-1}\right)^{a}=\sum_{t}
\sum_{n=1}^{|t|}\widetilde{N}(n,t)\frac{a(a-1)\cdots (a-n+1)}{n!}
\frac{X^{t}}{\mathrm{S}_{t}}.
\end{equation}
Then, by comparing with the expression of $(\mathrm{id}-X)^{-a}$ derived from \eqref{binomX}, we get
\begin{equation}
\sum_{n=1}^{|t|}\widetilde{N}(n,t)\frac{a(a-1)\cdots (a-n+1)}{n!}
=(-1)^{|t|}\sum_{n=d_{t}}^{|t|}N(n,t)(-1)^{n}\frac{a(a+1)\cdots (a+n+1)}{n!}
\end{equation}
By evaluating for various values of $a$, we get various relations between $N(n,t)$ and $\tilde{N}(n,t)$. For example, if $a=-1$ we obtain the sum rule
\begin{equation}
1=(-1)^{|t|}\sum_{n=d_{t}}^{|t|}(-1)^{n}N(n,t),
\end{equation}
which is a convenient way to check the computation of $N(n,t)$. For $a=-2$, we get
\begin{equation}
2+\widetilde{N}(2,t)=\sum_{n=d_{t}}^{|t|}(-1)^{n+|t|}(n+1)N(t,n),
\end{equation}  
which counts  the number of terms in the coproduct.

The geometric series and its cousins  can be used in two complementary ways. In many cases, the equations we encounter in physics can be written as fixed point equations of the type \eqref{fixedpoint}, so that rooted trees allow for an efficient determination of their perturbative solutions. For instance, this is the case for the self-consistency condition derived in mean field theory or the Schwiger-Dyson equations in QFT. In mathematics, this applies to the solution of an ordinary differential equation written in integral form, as we shall work out in section \ref{diffsec}, or to the Lagrange inversion formula.

Alternatively, one can be given a pertubative expansion exhibiting a natural tree like structure.  If one can find an operator $X$ such that this expansion can be written as a geometric series \eqref{geometrictree}, then this expansion is known to be a solution of the fixed point equation \eqref{fixedpoint}. Accordingly, the solution of this equation provides a good candidate for the sum of the tree like expansion.  We illustrate this procedure in the next section.

\subsection{Wigner's semi-circle law from rooted trees}

Let us give a very simple illustration of way one can use rooted trees to recover Wigner's semi-circle law of an ensemble of Gau\ss ian Hermitian $N\times N$ matrices. This technique is nothing but  a rooted tree formalisation of the treatment that can be found in \cite{zee}. 

We are interested in the large $N$ behavior of the density of eigenvalues, defined as
\begin{equation}
\rho(\lambda)=\lim_{N\rightarrow\infty}\int [DM]\,\mathrm{e}^{-N\mathrm{Tr}V(M)}
\sum_{i}\frac{\delta(\lambda-\lambda_{i})}{N}
\end{equation}
with $\lambda_{i}$ denoting the eigenvalues of $M$ and $V(M)=\frac{1}{2}M^{2}$. $[DM]=\prod_{i}dM_{ii}\prod_{i<j}d\mathrm{Re}M_{ij}d\mathrm{Im}M_{ij}$ is standard Lebesgue measure on the space of $N\times N$ Hermitian matrices and $\delta$ the Dirac distribution.

Using $\delta(x)=\frac{1}{\pi}\mathrm{Im}\left(\frac{1}{x-\mathrm{i}\epsilon}\right)$ with $\epsilon\rightarrow 0^{+}$, we can rewrite $\rho$ as
\begin{equation}
\rho(\lambda)=\lim_{N\rightarrow\infty}\frac{1}{N\pi}\mathrm{Im}\left\{
\int [DM]\mathrm{Tr}\left(\frac{1}{\lambda-\mathrm{i}\epsilon-M}
\right)
\mathrm{e}^{-N\mathrm{Tr}V(M)}\right\}.
\end{equation}
Introducing $z=\lambda-\mathrm{i}\epsilon$ to simplify the notation, we expand the resolvent as a geometrical series 
\begin{equation} 
\frac{1}{z-M}=\frac{1}{z}+\frac{1}{z}M\frac{1}{z}+\frac{1}{z}M\frac{1}{z}M\frac{1}{z}+\cdots.
\end{equation}

The expectation value $G(z)$ of  $\frac{1}{z-M}$ is easily computed using Wick's theorem with the Gau\ss ian matrix integral. It leads to Feynman rules with a single line propagator for $\frac{1}{z}$ a double line propagator $\frac{1}{N}$ for the propagation of the matrix field and a vertex set to 1 for the interaction of 2 single lines with a double line. Then, each face of the resulting diagram yields a factor $N$ because of the summation over matrix indices. Thus, the leading contribution in the large $N$ limit is given by those diagrams that have as many faces as possible for a fixed number of propagator. These diagrams are nothing but the celebrated planar diagrams.

The planar Feynman diagrams that contribute to $G(z)$ are in one to one correspondence with planar rooted trees (i.e. rooted trees counted as different if they differ by a permutation of their vertices).  The bijection is obtained by  drawing boxes that are either disjoint or nested around the matrix propagators, with the diagram always in the outermost box, even if there is no matrix propagator. This outermost box defines the root and the tree is drawn from the root to the leaves by following the natural hierarchal structure of the inclusion of the boxes.

\begin{figure}
\begin{center}
\includegraphics[width=15cm]{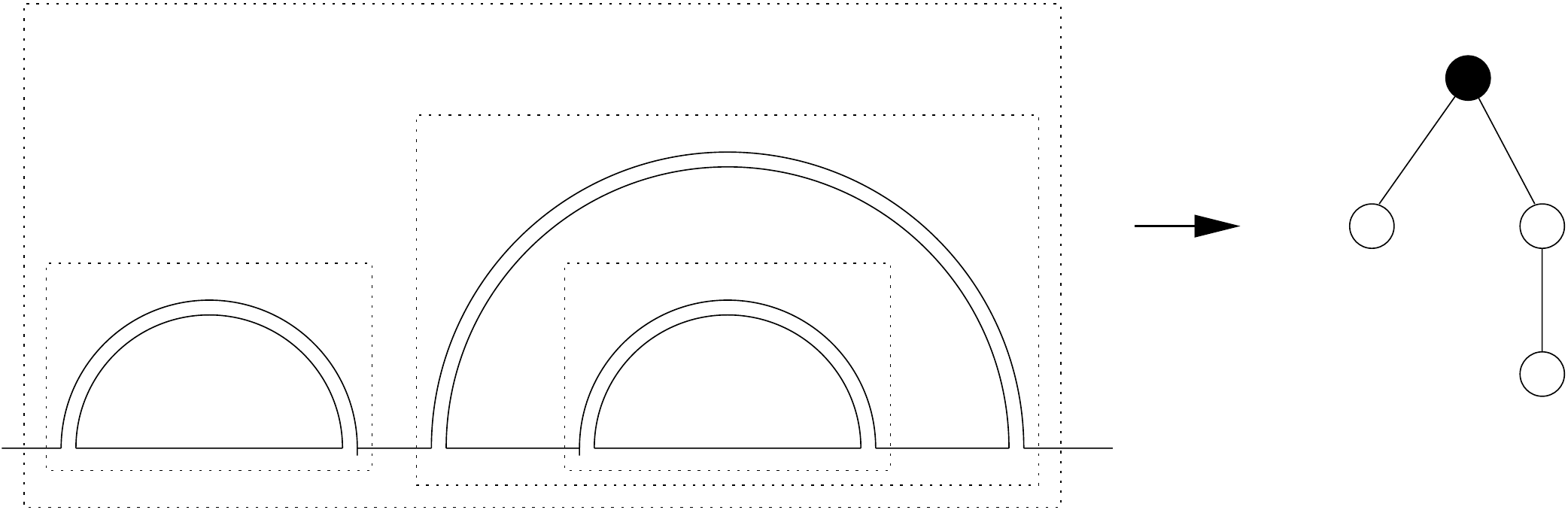}
\caption{A planar diagram contributing to $G(z)$}
\end{center}
\end{figure}

Isomorphic planar trees correspond to Feynman diagram having the same contribution to $G(z)$. Moreover, recall that a rooted tree is an isomorphism class of planar rooted tree and the sum over all representatives of the class of $t$ equals $\frac{X^{t}(0)}{\mathrm{S}(t)}$, with
\begin{equation}
X(x)=\frac{1}{z-x}.
\end{equation}
Note that the $n^{\mathrm{th}}$ order derivative of $X$ generates a factor $n!$ at each vertex with $n$ outgoing edges. When multiplied together along the tree and divided by the symmetry factor, the resulting number counts the cardinality of the isomorphism class of planar trees corresponding to $t$.

Therefore, $G(z)$ is the sum of the non linear geometric series
\begin{equation}
G(z)=\sum_{t}\frac{X^{t}(0)}{\mathrm{S}(t)}\,\mathrm{I_{N}},
\end{equation}
with $\mathrm{I_{N}}$ the $N\times N$ identity matrix. The sum of the geometric series is the solution $x$ of the quadratic fixed point equation $X(x)=x$ with leading term $\frac{1}{z}$, so that
\begin{equation}
G(z)=\frac{z-\sqrt{z^{2}-4}}{2}\,\mathrm{I_{N}}.
\end{equation}
and we recover Wigner's semicircle law as
\begin{equation}
\rho(\lambda)=\frac{1}{2\pi}\sqrt{4-\lambda^{2}},
\end{equation}
with support in $[-2,+2]$.

If we expand $G(z)$ in powers of $\frac{1}{z}$, we recover the ubiquitous Catalan numbers $C_{n}=\frac{(2n)!}{(n!)^{2}(n+1)}$,
\begin{equation}
G(z)=\sum_{n=0}^{\infty}C_{n}\frac{1}{z^{2n+1}}\,\mathrm{I_{N}}.
\end{equation}
This is not surprising since the Catalan number $C_{n}$ count, amongst many other things \cite{stanley}, the number of non crossing partitions of a set of $2n$ elements into blocks of 2. More sophisticated examples of summation over rooted trees applied to random matrices are described in the lecture by P. Di Francesco \cite{difrancesco}. In particular, it is shown how some similar techniques can be used for a quartic potential $V(M)=\frac{1}{2}M^{2}+\frac{g}{4}M^{4}$.

\subsection{Perturbative solution of differential equation}

\label{diffsec}

To solve pertubatively a time dependent non linear differential equation
\begin{equation}
\frac{dx}{ds}=X_{s}(x),\label{timede}
\end{equation}
with boundary condition $ x(s_{0})=x_{0}$, it is convenient to write it in integral form,
\begin{equation}
x(s)=x_{0}+\int_{s_{o}}^{s}ds'\, X_{s'}(x(s')).
\end{equation}
This is a fixed point equation for the integral non linear operator $x(s)\mapsto\int_{s_{o}}^{s}ds'\, X(x(s'))$ with the function $s\mapsto x(s)$ as the unknown. Using the geometric series, it solution is expanded over trees as
\begin{equation} 
x(s)=\sum_{t}\frac{1}{\mathrm{S}_{t}}\int_{I^{t}_{s,s_{0}}}\!\!d^{|t|}s\,
\prod_{v\in t}^{\longrightarrow}X_{s^{v}}^{(n_{v})}(x_{0}).
\end{equation}
To explain the meaning of these terms, recall that any tree is orientated form the root to the leaves. Then, associate to each vertex $v$ a real variable  $s_{v}$ in such a way that $s_{0}\leq s_{v}\leq s_{v_{+}}$ if $v_{+}$ is a vertex immediately above $v$  and $s_{v_{+}}=s$ for the root. This defines what we call a  "treeplex" $I^{t}_{s,s_{0}}\subset{\Bbb R}^{|t|}$, that reduces to a ordinary simplex if  $t$ is a  tree without branching. Then compose all the multilinear operators $X_{s^{v}}^{(n_{v})}(x_{0})$ from the root to the leaves and integrate over the variables $s_{v}$. Obviously, for linear $X$ we recover the ordinary time ordered products, with an integral over a simplex instead of a treeplex.

If we write the solution of \eqref{timede}, as 
$
\Omega_{s, s_{0}}(x_{0})=\sum_t\Omega^{t}_{s,s_0}(x_{0}), 
$
the semigroup property holds
\begin{equation}
\Omega_{s_{2},s_{1}}\circ\Omega_{s_{2},s_{1}}=\sum_{\mathrm{admissible}\,\mathrm{cuts}\,c}
\Omega^{t_{0}}_{s_{2},s_{1}}\circ_{c}
\left[\Omega^{t_{1}}_{s_{1},s_{0}},\dots,\Omega^{t_{n}}_{s_{1},s_{0}}\right],\label{semi}
\end{equation}
where $t_{0}$ is the part of $t$ that contains its root after the cut and $t_{1},\dots,t_{n}$ are the branches we cut. The symbol $\circ$ means composition followed by Taylor expansion and $\circ_{c}$ is the same operation, to be performed at each vertex immediately above the cut. This relation is similar to the definition of the coproduct and an analogous equation exists for the antipode, related to the inversion of $\Omega_{s,s_{0}}$.

\begin{figure}
\begin{center}
\includegraphics[width=8cm]{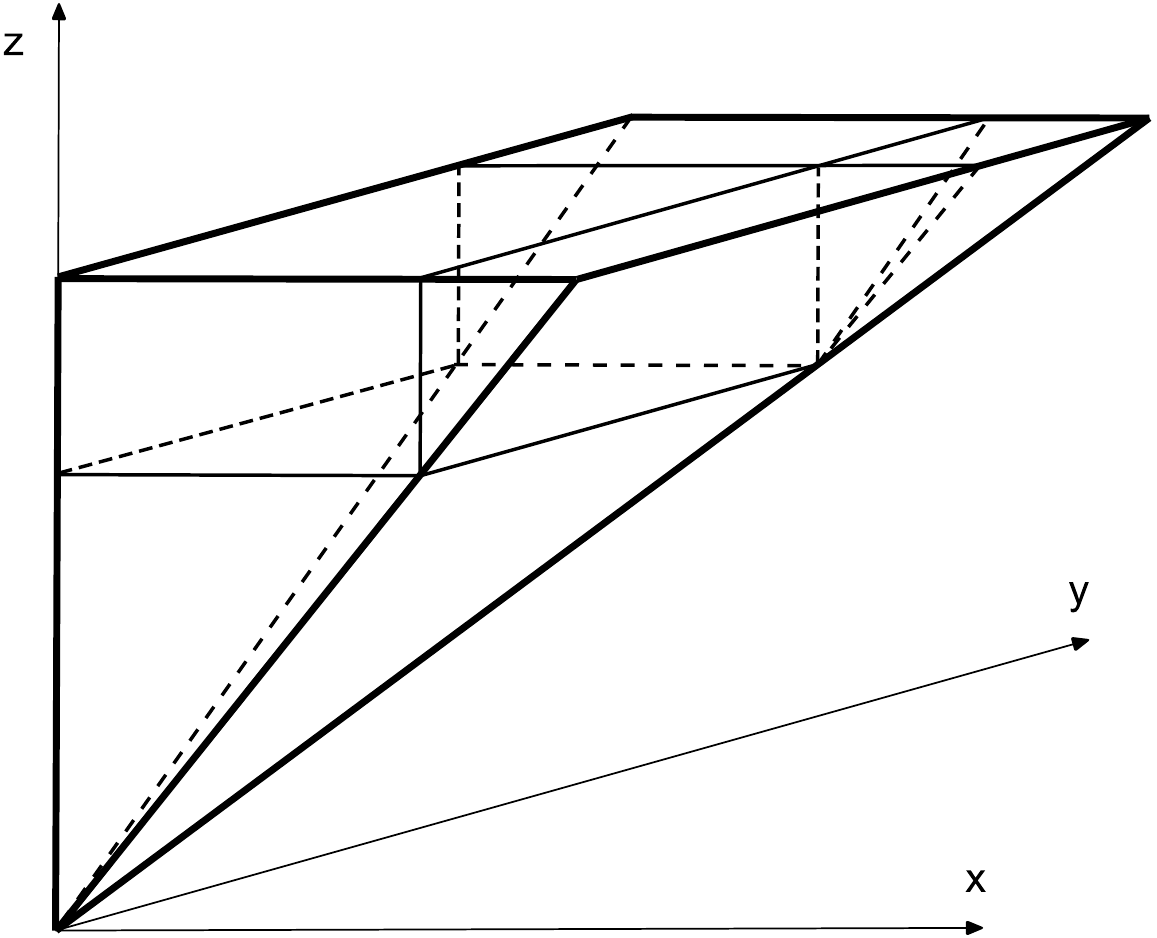}
\caption{Geometric interpretation of the coproduct}
\label{geometricinter}
\end{center}
\end{figure}

Comparing both sides of \eqref{semi} yields a geometric interpretation of the coproduct:
Cut $I_{s_{2},s_{0}}^{t}$ by a plane $s_{\mathrm{root}}=s_{1}$, so that it is written as a disjoint union
\begin{equation}
I_{s_{2},s_{0}}^{t}=\bigcup_{c\,\mbox{\tiny{admissible cut}}}{\frak
S}_{c}\left( I_{s_{1},s_{0}}^{t_{1}}\times\dots\times
I_{s_{1},s_{0}}^{t_{n}}\times I_{s_{2},s_{1}}^{t_{0}}\right)
\end{equation}
with $P_{c}(t)=t_{1}\dots t_{n}$ and $R_{c}(t)=t_{0}$. ${\frak
S}_{c}$ a suitable permutation of the labels preserving the ordering of the
tree. For example, for the simplest tree with one branching, this is illustrated on figure \ref{geometricinter} and the polytopes in this decomposition are in direct correspondence with the terms in \eqref{examplecop}.

For a time independent equation, the integral over $I^{t}_{s,s_{0}}$ factorizes and defines the tree factorial
\begin{equation}
\int_{I^{t}_{s,s_{0}}}d^{|t|}s=\frac{1}{t!}.
\end{equation}
The latter can be computed recursively: If $t$ is a tree obtained by grafting the trees $t_{1},\dots,t_{n}$ to the root, then
\begin{equation}
t!=\left(|t_{1}|+\cdots+|t_{n}|+1\right)\,t_{1}!\cdots t_{n}!.
\end{equation}
For a tree without branching we recover the ordinary factorial $t!=|t|!$.

Accordingly, the map $\Omega_{s,s_{0}}$ is a power series of the non linear operator $X$
\begin{equation}
\Omega_{s,s_{0}}=\sum_{t}\frac{1}{\mathrm{S}_{t}}\,\frac{ (s-s_{0})^{|t|}}{t!}\,X^{t},
\end{equation}
which is a non linear analogue of the exponential. The exponential character can be written as 
\begin{equation}
\exp_{\ast}\left\{(s-s_{0})\delta\right\}(t)=\frac{(s-s_{0})^{|t|}}{t!}\label{expo}
\end{equation}
with $\delta\in{\cal G}_{T}$ the infinitesimal character that takes the value 1 on the simplest non empty tree and vanishes otherwise. It satisfies the linear equation
\begin{equation}
\frac{d}{ds}\,\exp_{\ast}\left\{(s-s_{0})\delta\right\}=\delta\ast \exp_{\ast}\left\{(s-s_{0})\delta\right\}=\exp_{\ast}\left\{(s-s_{0})\delta\right\}\ast\delta.
\end{equation}
Thus, we have traded a non linear equation for a linear one, at the price of working with power series of non linear operators. 

The relation \eqref{expo} provides a combinatorial definition of the tree factorial, in analogy with what we did for the binomial series \eqref{binomial}. Because $\delta$ is an infinitesimal character, there are tremendous simplifications in the convolution powers of $\delta$ are evaluated on the iterated coproduct and only one term survives
\begin{equation}
\delta^{\ast n}(t)=N(n,t)
\end{equation}
if $n=|t|$ and vanishes otherwise. Therefore, $\frac{1}{t!}=\frac{N(|t|,t)}{|t|}$ so that
\begin{equation}
t!=\frac{\#\left\{ \mbox{labelings of the set of vertices in}\left\{1,\dots,n\right\} \right\}}
{\#\left\{ \mbox{labelings of the set of vertices decreasing from the root to the leaves} \right\}}.
\label{combifac}
\end{equation}

Turning back to numerical analysis, the Runge-Kutta method \eqref{RK} one naturally associates a  character $\alpha$ constructed as follows. 
Decorate the vertices of the tree $t$ by indices in $\left\{1,2,\dots,n\right\}$ and associate each edge from $i$ to $j$ with the matrix element $a_{ij}$. Multiply all these matrix elements together with a factor $b_{i}$ associated to the root and sum over all indices. Then, the approximate solution after one step reads 
\begin{equation}
x_{1}=\sum_{t}\frac{h^{|t|}}{{\mathrm S}_{t}}\,\alpha(t)\,X^{t}(x_{0}).
\end{equation}
It is a method of order $n$ if $\alpha$ coincides with the exponential on all trees of order at most $n$. This defines another formal power series of non linear operators and the composition law of Runge-Kutta methods is represented by the convolution of their characters.

\label{difftime}

\section{Renormalization, effective actions and Feynman diagrams}

\subsection{Renormalization for a differential equation}

\label{diffren}

Before dealing with path integrals and Feynman diagrams, it is fruitful to first investigate the perturbative renormalization for an ordinary differential equation \cite{GKM}. Recall that the differential renormalization group equation \eqref{RG} reads
\begin{equation}
\Lambda\frac{d
S_{\Lambda}}{d\Lambda}=\beta(\Lambda,S_\Lambda),\label{generalRGE}
\end{equation}
with boundary condition $S_{\Lambda_{0}}=S_{0}$ at a very high energy scale $\Lambda_{0}$.  In physics, $S_{\Lambda}$  is an effective  action that encodes the description of the physical system under scrutiny, valid up to an energy scale $\Lambda$, while  $S_{0}$ is the bare action which agrees with the effective action at the cut-off scale $\Lambda_{0}$. Ignoring temporarily the path integral interpretation of the renormalization group equation, let us simply consider a general equation of the form \eqref{generalRGE} for a variable $S$ in a Banach space ${\cal E}$. Following the general interpretation of renormalization, we are interested in the low energy behavior of $S_{\Lambda}$ for $\Lambda\ll \Lambda_{0}$. 

Because all the physical quantities can be measured in unit of mass\footnote{In the system of units $\hbar=c=1$ which we adopt here.}, we further assume that the space ${\cal E}$ comes equipped with an action of the dilatation group ${\Bbb R}^{\ast+}$. This action is  written in exponential form as  $\mathrm{e}^{s{\cal D}}$, with ${\cal D}$ a linear operator. It is often convenient to assume that ${\cal D}$ is diagonal, with eigenvalues corresponding to the canonical dimensions. Dimensional analysis is assumed to be compatible with the renormalisation group equation,
\begin{equation}
\beta(\mathrm{e}^{s}\Lambda,\mathrm{e}^{s{\cal D}}
S)=\mathrm{e}^{s{\cal D}}\beta(\Lambda,S).
\end{equation}
To take dimensional analysis into account, it is convenient to  introduce the dimensionless variables
$s=\log(\Lambda/\Lambda_{\mbox{\tiny ref}})$ with
$\Lambda_{\mbox{\tiny ref}}$ a fixed reference scale,
$u(s)=\Lambda^{-{\cal D}}\!\cdot\!S_{\Lambda}$ and
$X(u)={\Lambda}^{-{\cal D}}_{\mbox{\tiny
ref}}\!\cdot\!\beta({\Lambda}_{\mbox{\tiny ref}},{\cal
S}_{\Lambda})$. In terms of these new variables, the renormalization group differential equation takes the simpler form
\begin{equation}
\frac{du}{ds}=-{\cal D}u+X(u),
\end{equation}
with boundary condition $u(s_{0})=u_{0}$. 

The pertubative solution of this equation is easily obtained by writing it in integral form
\begin{equation}
u(s)=\mathrm{e}^{-(s-s_{0}){\cal D}}u_{0}+\int_{s_{o}}^{s}ds'\, \mathrm{e}^{(s-s'){\cal D}}\,X(u(s')).\label{fixedrg}
\end{equation}
Making use of the geometric series of the integral operator ${\cal F}\cdot u(s)=\int_{s_{o}}^{s}ds'\, \mathrm{e}^{(s-s'){\cal D}}\,X(u(s'))$, 
\begin{equation}
u=\left(\mathrm{id}-{\cal F}\right)^{-1}(u_{0})\label{diffpert}
\end{equation}
we expand $u(s)$ as a sum over trees,  using the following rules to compute the contribution of each tree $t$. First, orient all the edges from the root to the leaves and then associate 

\begin{itemize}

\item
variables $s_{v}$ to the vertices that leave in the the treeplex $I_{s_{0},s}^{t}$.

\item
linear operators $\mathrm{e}^{-(s_{v_{+}}-s_{v_{-}}){\cal D}}$ to edges from $v_{+}$ to $v_{-}$ (recall that edges are orented from the root to the leaves); 

\item multilinear operators $X^{(n_{v})}(\mathrm{e}^{-(s_{v}-s_{0}){\cal D}}  u_{0})$
 to vertices with $n_{v}$ outgoing edges.

\end{itemize}

Finally, compose from the root to the leaves all operators
associated to edges and vertices, multiply on the left by $\mathrm{e}^{-(s-s_{\mathrm{root}})\delta}$ and integrate over the variables associated to the vertices. This is very similar to the expansion we encountered in the previous section for a time dependent equation, at the notable exception of the insertion of the of the linear operators  $\mathrm{e}^{-(s_{v_{+}}-s_{v_{-}}){\cal D}}$ along the edges.

To simplify the notations, let us abbreviate as $X_{s_{v}}^{(n)}$ the $n$-linear map obtained by differentiating $X$ $n$ times at $\mathrm{e}^{-(s_{v}-s_{0}){\cal D}}  u_{0}$. For example, the linear tree of order two contributes as
\begin{equation}
\int_{s_{0}}^{s}\! ds_{2}\int_{s_{0}}^{s_{2}}\! ds_{1} \quad
\mathrm{e}^{-(s-s_{2}){\cal D}}
X'_{s_{2}}
\Big[\mathrm{e}^{-(s_{2}-s_{1}){\cal D}} X_{s_{1}}\Big],
\end{equation}
whereas the order 3 tree with one branching yields
\begin{equation}
\frac{1}{2}
\int_{s_{0}}^{s}\! ds_{3}\int_{s_{0}}^{s_{3}}\! ds_{2} 
\int_{s_{0}}^{s_{3}}\! ds_{1} 
\quad
\mathrm{e}^{-(s-s_{3}){\cal D}}
X''_{s_{3}}
\Big[\mathrm{e}^{-(s_{3}-s_{2}){\cal D}} X_{s_{2}},\,
\mathrm{e}^{-(s_{3}-s_{1}){\cal D}} X_{s_{1}}
\Big]
\end{equation}
In the simplest case of a linear $X$, the contributions of all branched trees vanish and we recover Duhamel's formula (or in the physics terminology, the Schwinger-de Witt expansion) of the exponential of $-{\cal D}+X$ in powers of $X$,

In analogy with perturbative renormalization in QFT, let us try to get rid of the cut-off by taking the limit $s_{0}\rightarrow\infty$. If we discard all the non linear terms, the analysis us easy to perform: 

\begin{itemize}
\item
variables $u^{\mathrm{irr}}$ that belong to the negative eigenspaces of ${\cal D}$ (irrelevant variables) tend to $0$;
\item
variables $u^{\mathrm{mar}}$ that belong to the zero eigenstpace of ${\cal D}$ (marginal variables) remain constant;
\item
variables $u^{\mathrm{rel}}$ that belong to the positive eigenspaces of ${\cal D}$ (relevant variables) grow to infinity.
\end{itemize}

In perturbation theory, this picture is slightly modified by the addition of the non linear terms. Marginal variables have a polynomial dependence on $s$, analogous to the $\log\Lambda$ terms in QFT. Moreover, irrelevant terms at a given order depend on the marginal and relevant ones at a lower order, so that they also experience divergences. This phenomena is analogous to the subdivergences of QFT.

The solution of the problem of UV divergencies is natural from the Wilsonian viewpoint: Instead of imposing the boundary condition for the divergent variables at the very high energy $s_{0}$ tending to infinity, impose the latter at a fixed low energy $s_{\mathrm{r}}$. This corresponds to a determination of the relevant (and marginal) parameters by a measurement at low energy. To implement this change of boundary condition, let us denote by ${\cal P}$ the projector onto the relevant and marginal variables and by ${\cal F}'$ the non linear operator  
\begin{equation}
{\cal F}'[u](s)={\cal P}
\int_{s_{\mathrm{r}}}^{s_{0}}ds'\, \mathrm{e}^{(s-s'){\cal D}}\,X(u(s')).
\end{equation}
Therefore, the renormalized theory, expressed as a function of the renormalized parameters (irrelevant paramaters at $s_{0}$, relevant and marginal ones at $s_{\mathrm{r}}$), is obtained as a fixed point for the renormalized operator ${\cal F}+{\cal F}'$. To compare the renormalized and unrenormalized expansions, let us write
\begin{equation}
\Big[\mathrm{id}-\left({\cal F}+{\cal F}'\right)\Big]^{-1}= \Big[\mathrm{id}-{\cal F}\Big]^{-1}\circ
\Big[\mathrm{id}-{\cal F}'\circ\left(\mathrm{id}-{\cal F}\right)^{-1}\Big]^{-1}.
\end{equation}
In the QFT language, the LHS is the renormalized theory while the RHS is the unrenormalized one composed with the counterterms. The latter are the boundary conditions on the relevant and marginal parameters to be imposed at the high energy $s_{0}$ in such a way that these parameters match their fixed values at low energy $s_{\mathrm{r}}$. At the level of the tree expansions, the expansion of the renormalized theory can be pictured as a sum over trees with  both white vertices for ${\cal F}$ and black ones for ${\cal F}'$, while the unrenormalized theory only involves white vertices.
Counterterms only involve trees with a black root, followed by black and white vertices that correspond to the renormalization of the subdivergences. This is the analogue of the BPHZ formula for a differential equation.  

In the renormalized expansion, all the $s_{0}$ dependence has dropped from the divergent exponentials and their integrals,  so that the contribution of each tree admits a finite limit as $s_{0}\rightarrow\infty$.. Moreover, the initial conditions $u_{0}^{\mathrm{irr}}$ for the irrelevant paremeters all enter through the combination $\mathrm{e}^{-{\cal D}(s-s_{0})}u_{0}^{\mathrm{irr}}$, which goes to $0$ as $s_{0}\rightarrow\infty$.  In this way we recover universality: The renormalized theory does not dependent on the choice of the irrelevant parameters at high energy. 
 
When formulated in terms of differential equations, the renormalization procedure is general enough to encompass Polchinski's equation \eqref{erge}, as far as perturbation is concerned. Solving the latter using rooted trees yields a smart of proof of the renormalizability of $\phi^{4}$ theory in $D=4$ that does not refer to Feynman diagrams \cite{Hurd}, in addition to the original proof by Polchinski \cite{Polchinski}.

Though rooted trees are a convenient device that often leads to convergent expansions, it is fair to say that this is seldom the case for differential equation with boundary condition at $\infty$.  To illustrate the issue raised by non perturbative computations within the differential renormalization group approach, let us consider a very simple example with one marginal and one irrelevant coupling,
\begin{equation}
\renewcommand{\baselinestretch}{3}
\left\{
\begin{array}{rclr}
\renewcommand{\baselinestretch}{3}
\frac{dg}{ds}&=&\beta g^{2}&({\cal D}=0),\cr \frac{du}{ds}&=&u+\gamma
g&({\cal D}=-1).\label{toybeta}
\end{array}
\right.
\end{equation}
\renewcommand{\baselinestretch}{2}
This system can be solved using the pertubative techniques we have presented. However, because of the extreme simplicity of these equations, it is much more convenient to determine an exact solution. Let us first investigate the case of the marginal coupling constant $g$. In the unrenormalized case, the boundary condition is imposed at very high energy, so that the unrenormalized solution and its expansion read
 \begin{equation}
g(s)=\frac{g_{0}}{1-\beta g_{0}(s-s_{0})}=g_{0}\sum_{n=0}^{\infty}(\beta g_{0})^{n}(s-s_{0})^{n}.
\label{bareg}
 \end{equation}
In terms of rooted trees, the LHS is the sum over rooted trees with vertices having only at most two   outgoing edges (since the operator $g\mapsto\beta g^{2}$ is quadratic) and a weight corresponding to the tree factorial.

As expected, the coefficients of the expansion, which play the role of the Feynman diagrams,  diverge as $s_{0}\rightarrow\infty$ at fixed $g_{0}$. Perturbative renormalization is achieved by imposing boundary conditions at a fixed low energy scale $g(s_{\mathrm{r}})=g_{\mathrm{r}}$, so that the renormalization coupling constant reads
 \begin{equation}
g(s)=\frac{g_{\mathrm{r}}}{1-\beta g_{\mathrm{r}}(s-s_{\mathrm{r}})}=g_{\mathrm{r}}\sum_{n=0}^{\infty}(\beta g_{\mathrm{r}})^{n}(s-s_{\mathrm{r}})^{n}.
\label{reng}
\end{equation}
The transition from the unrenormalized coupling to renormalized one is achieved by choosing $g_{0}$ in such a way that $g$ obeys the new boundary condition $g(s_{\mathrm{r}})=g_{\mathrm{r}}$, which is easily seen to be
 \begin{equation}
g_{0}=\frac{g_{\mathrm{r}}}{1-\beta g_{\mathrm{r}}(s_{0}-s_{\mathrm{r}})}=g_{\mathrm{r}}\sum_{n=0}^{\infty}(\beta g_{\mathrm{r}})^{n}(s_{\mathrm{r}}-s_{0})^{n}.
\end{equation}
Then, as we substitute this expresssion of $g_{0}$ in \eqref{bareg} we recover \eqref{reng}. At the perturbative level, this substitution is best seen using rooted trees and illustrates the way BPHZ renormalization must be performed with the trees.

\begin{figure}
\begin{center}
\includegraphics[width=12cm]{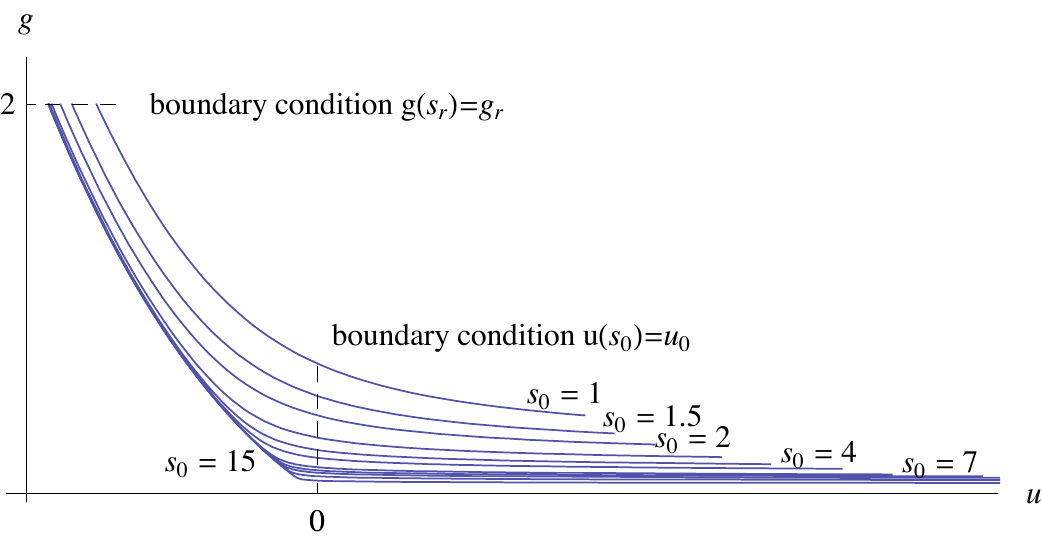}
\caption{Convergence towards the renormalized theory  for the differential equation \eqref{toybeta}}
\end{center}
\end{figure}

Once the renormalized coupling constant \eqref{reng} is known, the solution of the second equation follows from the method of variation of constants,
\begin{equation}
u(s)=u_{0}\mathrm{e}^{s-s_{0}}+\int_{s_{0}}^{s}\,ds'
\frac{\gamma g_{\mathrm{r}}\,e^{-(s'-s)}}{1-\beta g_{\mathrm{r}}(s'-s_{r})}.\label{exact}
\end{equation}
First, let us notice that the $u$ is independent of $u_{0}$ as $s_{0}\rightarrow\infty$. This is universality in this simple context: The renormalized theory does not depend on the irrelevant terms at high energy. 
At $s_{\mathrm{r}}=0$ and $s_{0}\rightarrow\infty$, the exact solution can be expanded as
\begin{equation}
u(g_{\mathrm{r}})=-\gamma\int_{0}^{\infty}\!ds\,e^{-s}\frac{g_{\mathrm{r}}}{1-g_{\mathrm{r}}\beta
s}\simeq
-\mathop{\sum}\limits_{n=0}^{\infty}\beta^{n}n!(g_{\mathrm{r}})^{n+1}.
\end{equation}
Though it provides accurate approximations of $u$ for small $g_{\mathrm r}$, it is a divergent power series, Borel summable only if $\beta<0$. Nevertheless, if $\beta<0$ the coupling $g(s)$ remains bounded and $u$ is well defined as a functional of $g:\,s\mapsto g(s)$,
\begin{equation}
u[g]=\int_{0}^{\infty}ds\,e^{-s}\,g(s)
\end{equation}
This is a particularly simple example of an effective expansion \cite{constructive}, whose counterpart in QFT is the  fact that a asymptotically free theory, characterized by $\beta<0$, may be well defined when expanded in a scale dependent coupling instead $g(s)$ of a single renormalized coupling $g_{\mathrm{r}}$.  In case $\beta>0$, \eqref{exact} can still be given a mathematical meaning by deforming the contour of integration in the $s$ plane. However, in this case there are extra ambiguities related to the monodromy around the pole so that the construction of the theory requires extra information not contained in the perturbation theory.  Nevertheless, the theory may lose its physical meaning because it is no longer real.

The occurrence of the previous type of $n!$ behavior that prevents a naive summation of the perturbative solution with boundary conditions at infinity is due to the phenomenon of resonances. Consider a general renormalization group equation  
\begin{equation}
\frac{du}{dt}=-{\cal D}u+X(u)\label{non linear}
\end{equation}
and assume that ${\cal D}$ has been diagonalized in a basis $(e_{i})$ with eigenvalues $\lambda_{i}$. We further assume that $X$ can be written as a polynomial (or an analytic function)
\begin{equation}
X(u)=\sum_{i,j_{1}\dots j_{n}}e_{i} X^{i}_{j_{1}\dots j_{n}}
(u_{j_{1}})^{m_{1}}\cdots (u_{j_{n}})^{m_{n}}.\label{decres}
\end{equation}
A monomial in \eqref{decres} is resonant if there is a non trivial relation of the type
\begin{equation}
\lambda_{i}=m_{1}\lambda_{j_{1}}+\cdots+m_{n}\lambda_{j_{n}},\label{resonance}
\end{equation}
with $m_{k}\in{\Bbb N}$. 

As we try to solve the renormalization group equation pertubatively in $u$, the Poincar\'e-Dulac theorem states that, by a series of successive changes of variables of the type $v=u+F(u)$ with $F$ having at least a second order term, the renormalization group equation takes  the simpler form
\begin{equation}
\frac{dv}{dt}=-{\cal D}v+Y(v)\label{non linear}
\end{equation}
with only resonant monomials in $Y$. Because the resonance condition \eqref{resonance} can be formulated as 
\begin{equation}
\mathrm{e}^{-s{\cal D}}\circ Y=Y\circ\mathrm{e}^{-s{\cal D}},
\end{equation}
all the exponentials but the the first one cancel in the perturbative expansion of the solution given in \eqref{diffpert}. Therefore, the iterated integral along the tree generates a power law $s^{n}$ and leads to a factorial behavior $n!$ when integrated together with the exponential along the root.

Let us conclude this section by two remarks that may be the subject of some future work.

\begin{itemize}
\item
It could be interesting to derive the  normal form of the QFT renormalization group equation \eqref{erge}, with a view towards a non perturbative analysis. This could be done along the lines initiated in \cite{lebellac} for a toy model.
\item
Reduction to  normal forms is a non linear generalization of the Lie algebra decomposition into commuting semi-simple and nilpotent elements \cite{arnold}. As such, it could be conveniently described in the algebraic framework presented in the sections devoted to rooted trees and Feynman diagrams.

\end{itemize}

\subsection{Hopf algebra of Feynman diagrams and effective actions}

In the view of the algebraic structure we have presented in the previous sections, let us investigate Wilsonian renormalization in QFT. The basic renormalization group transform is the integration over fast modes that induces the transformation of effective actions $S\rightarrow S'$,
\begin{equation}
\mathrm{e}^{S'[\phi]}={\cal N}
\int[D\chi]\,\mathrm{e}^{-\frac{1}{2}\chi\cdot A^{-1}\cdot \chi}\,
\,\mathrm{e}^{S[\phi+\chi]}.
\end{equation}
$A$ is a suitable propagator that implements the integration over momenta lying between $\Lambda$ and $\Lambda_{0}$. In what followos, we consider $A$ to be arbitrary propagator, with suitable UV and IR cut-offs. ${\cal N}$ is a normalization factor, independent of $S$ and $\phi$, that ensures that $S'=0$ for $S=0$. In the sequel, we do not display such normalization factors anymore. Note that we have used $S$ instead of $-S$ in the exponential to avoid some irrelevant signs.

The renormalization group transformation of  effective actions is best formulated using the background field method. The rationale of the background field method is to split the field into a fixed classical field $\phi$ and a quantum field $\chi$ over which we integrate, so that we end up with an effective action function for the classical field $\phi$. In our setting, the fast modes play the role of the quantum field whereas the low energy ones remain as an argument of the low energy effective action. Using this technique, $S'$ is expanded over connected vacuum Feynman diagrams (i.e. diagrams without external leg) as
\begin{equation}
S'[\phi]=\sum_{\gamma}\frac{A^{\gamma}(S)}{\mbox{
S}_{\gamma}}[\phi].
\end{equation}
The sum starts with the trivial empty diagram whose contribution is $\cal{S}[ \phi]$. For other diagrams, $A^{\gamma}(\cal S)$ is computed using the background field
technique Feynman rules: Propagators are determined by $A$ and $n$-valent vertices are $n^{\mathrm{th}}$ order functional differentials of $S$ at $\phi$.
As is always the case for diagrammatic expansions, one has to further divide by the symmetry factor $\mathrm{S}_{\gamma}$, which is nothing but the cardinal of the automorphism group of the diagram. All the dependence on $\phi$ is encoded in the vertices so that there is no need for external sources and we restrict ourselves to vacuum diagrams. Note that there is no restriction on $S$  since the convergence of each Feynman diagram is ensured by the cut-off.  Strictly speaking, this true only for a sharp cut-off, when considering a cut-off with a exponential decrease we have to assume that the momentum dependence of the functional derivatives of $S$ does not grow faster than a polynomial.

To illustrate the background field technique, let us give a simple example with a finite dimensional integral and a positive definite symmetric matrix $A_{ij}$ as a propagator.

\[
\xy
(0,0)*{}="A"; 
(16,0)*{}="B"; 
"A"; "B" **\crv{(0,0)&(8,8)&(16,0)};
"A"; "B" **\crv{(0,0)&(16,0)};
"A"; "B" **\crv{(0,0)&(8,-8)&(16,0)};
\endxy
\quad\rightarrow\quad
\frac{1}{12}
\sum_{i_{1},i_{2},i_{3}\atop
j_{1},j_{2},j_{3}}
\frac{\partial^{3}S}{\partial\chi^{i_{1}}\partial\chi^{i_{2}}\partial\chi^{i_{3}}}(\phi)\,
A_{i_{1},\,j_{1}}A_{i_{2},\,j_{2}}A_{i_{3},\,j_{3}}\,
\frac{\partial^{3}S}{\partial \chi^{j_{1}}\partial\chi^{j_{2}}\chi^{j_{3}}}(\phi)
\]

In the spirit of Wilsonian renormalization, one has to pursue this procedure and consider $S'$ to be a starting point for a new integration over yet slower modes. This is the reason why we are forced to deal with actions of a very general form. Indeed, even if $S$ is a simple $\phi^{4}$ interaction, this is no longer the case for $S'$ that contains arbitrarily complicated vertices. To handle this type of computation, it is necessary to modify the Hopf algebra defined by Connes and Kreimer as follows. Let ${\cal H}_{\mathrm{F}}$ be the free commutative algebra generated by all connected vacuum Feynman diagrams with vertices of arbitrary valence, including univalent and bivalent ones.  It is equipped with a Hopf algebra structure with trivial counit and coproduct
\begin{equation}
\Delta(\gamma)=\gamma\otimes 1+1\otimes\gamma+
\mathop{\sum}\limits_{\gamma_{1},\dots,\gamma_{n}\atop
\mathrm{disjoint\,subdiagrams}}
\gamma_{1}\cdots\gamma_{n}\otimes\gamma/(\gamma_{1}\cdots\gamma_{n}),
\label{coproductfeyn}
\end{equation}
where the sum runs over all possible mutually disjoint connected subdiagrams. In our setting, a subdiagram is a subset lines and vertices related to them. The reduced diagram $\Gamma/(\gamma_{1}\cdots\gamma_{n})$ is obtained by shrinking each $\gamma_{i}$ to a single vertex in $\Gamma$. Obviously, ${\cal H}_{\mathrm{F}}$ is graded by the number of internal lines $I$ but not by the number of vertices. However, the number of loops $L$ also provides a grading because of the identity $L=I-V+1$ for a connected diagram, but this grading does not yield a connected algebra since all the trees have grading 0. To these gradings is associated the two parameter group of automorphisms
\begin{equation}
\varphi_{a,b}(\gamma)=a^{I_{\gamma}}b^{L_{\gamma}}\,\gamma,\label{auto}
\end{equation}
with $I_{\gamma}$ (resp. $L_{\gamma}$) being the number of internal lines (resp. loops) of $\gamma$. Because of the grading, one can apply the general theory of graded and commutative Hopf algebras: The antipode may be defined recursively and characters of ${\cal H}_{F}$  form a group $G_{F}$, with a Lie algebra ${\cal G}_{F}$ of infinitesimal characters. Note that elements of both $G_{F}$ and ${\cal G}_{F}$ are uniquely defined by their values on the diagrams, which may assume any values in a given commutative algebra. As for trees, the exponential defines a map $\exp_{\ast}:\,{\cal G}_{F}\rightarrow G_{F}$ by
\begin{equation}
\exp_{\ast}(\delta)=\epsilon+\delta+\frac{\delta\ast\delta}{2}+\cdots+
\frac{\delta^{\ast n}}{n!}+\cdots,
\end{equation}
which is a solution of the differential equation
\begin{equation}
\frac{d}{ds}\,\exp_{\ast}(s\delta)=\delta\ast \exp_{\ast}(s\delta)=\exp_{\ast}(s\delta)\ast\delta
\label{diffFeyn}
\end{equation}
reducing to the unit of $G_{T}$ at $s=0$.

The elements of $G_{F}$ can be considered as weights of the for the Feynman diagrams in the expansion of the transformation of effective actions $\Psi^{A}_{\alpha}:\,S\rightarrow{\cal
S}'$
\begin{equation}
S'=\Psi^{A}_{\alpha}(S)=\sum_{\gamma}\alpha(\gamma)\frac{A^{\gamma}({\cal
S})}{{\mbox{{\small S}}_{\gamma}}}
\end{equation}
Given any two characters $\alpha,\beta\in G_{F}$, there is a composition law analogous to the one we encounter for power series of non linear operators in \eqref{composition},
\begin{equation}
\underbrace{\Psi^{A}_{\beta}\circ\Psi^{A}_{\alpha}}_{\mbox{\scriptsize
composition}}=
\underbrace{\Psi^{A}_{\alpha\,\ast\,\beta}}_{\mbox{\scriptsize
convolution}}.
\label{composeeff}
\end{equation}
This composition law describes how the weight of the Feynman diagrams entering in an iteration of two  effective action computations can be obtained by by inserting diagrams with weight $\alpha$ into diagrams with weight $\beta$. To prove this relation, we introduce $\delta=\log_{\ast}(\alpha)$ and $\alpha_{s}=\exp_{\ast}(s\delta)$. At first order in $s$, this relation does not involve multiple insertions of diagrams and is easily verified by simple combinatorics. Then, we integrate the differential equation to recover \eqref{composeeff}. In a certain sense, it is a Wilsonian analogue of the morphism from the group of graphs into the diffeomorphisms of the coupling constant presented in \cite{ck2}. 

In physics, there are many characters  that appear naturally. First, one may consider $\alpha$ to be the character that selects 1PI diagrams and $\beta$ the one the selects the trees. Then, $\alpha\ast\beta$ takes the value one on all diagrams and the composition law \eqref{composeeff} simply states that the one recovers the complete effective action by first computing 1PI diagrams and then inserting into trees. As another simple example, let us consider the character $\alpha_{s,h}(\gamma)=s^{I_{\gamma}}h^{l_{\gamma}}$, so that the corresponding effective action transformation $S\rightarrow S'$ can be expressed as
\begin{equation}
S'[\phi]=h\log\left\{
\int[D\chi]\,\mathrm{e}^{-\frac{1}{2}\chi\cdot (hsA)^{-1}\cdot \chi}\,
\,\mathrm{e}^{\frac{S[\phi+\chi]}{h}}
\right\}.
\label{loopeff}
\end{equation} 
As we expand $S'$ in term of Feynman diagrams, we find a weight $hs$ for every internal line and $h^{-1}$ for every vertex together with a global factor of $h$. Thus a connected diagram with $I_{\gamma}$ internal lines, $V_{\gamma}$ vertices and $L_{\gamma}=I_{\gamma}-V_{\gamma}+1$ loops is weighted by $s^{I_{\gamma}}h^{I_{\gamma}-V_{\gamma}+1}=s^{I_{\gamma}}h^{L_{\gamma}}.$ 
Other characters of interest are the polynomials in $N$ that arise in the $O(N)$ invariant models.

To handle more complicated situations, it may necessary to decorate the Feynman diagrams. For instance, if we want to perform the second effective action with a different propagator $B$, then we have to decorate the internal lines with a index that differentiates $A$ and $B$. With decorations on the internal lines, the Hopf algebra structure remains similar with the subgraph that we extract in \eqref{coproductfeyn} carrying the decorations. Then, we can perform a first effective action computation with propagator $A$ and character $\alpha$, followed by a second one with $B$ and $\beta$. The results amounts to a single computation with decorated internal lines that selects the type of propagators we use. Then, the analogue of \eqref{composeeff} reads
\begin{equation}
\Psi_{\beta}^{B}\circ\Psi_{\alpha}^{A}=\Psi_{\alpha\ast\beta}^{A,B}\label{AB},
\end{equation}
Both $\alpha$ and $\beta$ are characters of the Hopf algebra with two decorations on the internal lines, but $\alpha$ and (resp. $\beta$) vanish on diagrams that do not contain $A$ (resp. $B$). The previous relation is a generalization of the convolution of Gau\ss ian integrals
\begin{eqnarray}
\int[D\chi]\,\int[D\xi]\,
\exp\left\{-{\textstyle\frac{1}{2}}\chi A^{-1}\chi-{\textstyle \frac{1}{2}}\xi B^{-1}\xi+\-S[\phi+\chi+\xi]\right\}&=&\cr
\int[D\zeta]\,\exp\left\{-{\textstyle\frac{1}{2}}\zeta(A+B)^{-1}\zeta+{\cal
S}[\phi+\zeta]\right\},&&
\end{eqnarray}
which we obtain from \eqref{AB} when the characters $\alpha$ and $\beta$ only take the values 0 or 1, depending on whether the diagrams only involve the corresponding propagator. Obviously, this construction can be generalized to an arbitrary number of propagators so that we can iterate the effective action construction. Unfortunately, this construction is not sufficient to handle Polchinski's equation since the latter involve an infinite number of iterations corresponding to the differential equation. Moreover, the various types interactions (relevant, marginal and irrelevant) play fundamentally different roles so that it is necessary to distinguish them in the Feynman expansion. We postpone these questions to the section \ref{polchsec}, devoted to the iterative solution of Polchinski's equation.  Before we deal with the latter, let us  investigate some algebraic properties of the ${\cal H}_{F}$ and illustrate them in the next section on some properties of the Tutte polynomial.

First of all, let us notice that there are only two diagrams of the degree one (i.e. with one internal line): The bridge  with two vertices and the self-loop with a single vertex. Let us define two inifinitesimal characters: $\delta_{\mathrm{tree}}$ takes the value one on the bridge and vanishes otherwise, and $\delta_{\mathrm{loop}}$ takes the value one on the self-loop and vanishes otherwise,
\begin{equation}
\left\{
\begin{array}{lclcc}
\delta_{\mathrm{tree}}(\parbox{0.4cm}{\mbox{\includegraphics[width=0.4cm]{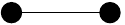}}})&=&1&&\cr
\delta_{\mathrm{tree}}(\gamma)&=&0&\mathrm{if}&\gamma\neq\parbox{0.4cm}{\mbox{\includegraphics[width=0.4cm]{tutte8.eps}}}
\end{array}
\right.
\qquad\mathrm{and}\qquad
\left\{
\begin{array}{lclcc}
\delta_{\mathrm{loop}}(\parbox{0.4cm}{\mbox{\includegraphics[width=0.4cm]{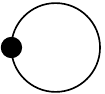}}})&=&1&&\cr
\delta_{\mathrm{loop}}(\gamma)&=&0&\mathrm{if}&\gamma\neq\parbox{0.4cm}{\mbox{\includegraphics[width=0.4cm]{tutte5.eps}}}
\end{array}
\right.
\end{equation}
By iterating the coproduct, it is easy to see that
\begin{equation}
\exp_{\ast}\left\{a\,\delta_{\mathrm{tree}}+b\,\delta_{\mathrm{loop}}\right\}(\gamma)=a^{I_{\gamma}\!-\!L_{\gamma}}b^{L_{\gamma}},
\label{diagexpo}
\end{equation}
where $I_{\gamma}$ is the number of internal lines and $L_{\gamma}$ the number of loops. By coassociativity, the iteration can be performed either on the left or on the right. If we perform it on the left, then we contract the internal lines and we get a factor of $a$ if the line is a self-loop or $b$ if it is not. When performed  on the right we delete internal lines and we also get factors of $a$ and $b$ depending on whether the line is a bridge (i.e. a line whose removal disconnects the diagram) or not. Note that all contraction/deletion schemes yield the same result and can be mixed together: At one step we can delete and then contract at another.

Any infinitesimal character $\delta\in{\cal G}_{F}$ defines two derivations of ${\cal H}_{F}$ that correspond to the infinitesimal form of right and left multiplication by $\exp_{\ast}(s\delta)$ on functions over $G_{T}$ at first order in $s$ ,
\begin{equation}
\left\{
\begin{array}{rcl}
f\lhd\delta&=&\left(\delta\otimes\mathrm{Id}\right)\circ\Delta(f),\cr
\delta\rhd f&=&\left(\mathrm{Id}\otimes\delta\right)\circ\Delta(f),
\end{array}
\right.
\end{equation}
for any $f\in{\cal H}_{T}$. Obviously, left and right action for different characters commute. In case of the infinitesimal characters $\delta_{\mathrm{tree}}$ and $\delta_{\mathrm{loop}}$, these operations have natural interpretations in terms of deletions and contraction of edges.  Recall that a self-loop is an edge with both ends linked to the same vertex and a bridge is an edge whose removal disconnects the diagram. Note that an edge connecting a single vertex to the rest of the diagram is considered as a bridge.
\begin{itemize}
\item
$\delta_{\mathrm{loop}}\rhd\gamma$ is a sum over all the diagrams obtained from $\gamma$ by cutting a line which is not a bridge. 
\item
$\delta_{\mathrm{tree}}\rhd\gamma$ is a sum over all the diagrams obtained from $\gamma$ by cutting 
a bridge.
\item
$\gamma\lhd\delta_{\mathrm{loop}}$ is a sum over all the diagrams obtained from $\gamma$ by contracting a self-loop with one edge.
\item
$\gamma\lhd\delta_{\mathrm{tree}}$ is a sum over all the diagrams obtained from $\gamma$ by contracting a line which is not a self-loop.
\end{itemize} 

As will be illustrated in the next section, these rules allow to translate simple diagrammatic manipulations into differential equations.  
 
 \label{diffrelations}

\subsection{An interlude on the Tutte polynomial}
 
The Tutte polynomial is a two variable polynomial associated to graphs that plays a fundamental role in graph theory and statistical mechanics (see \cite{tutte} for a recent review). In what follows, we shall illustrate its universality property in graph theory and its relation to the Potts model partition function using the techniques we introduced in the previous section.

The Tutte polynomial  is a polynomial in $x$ and $y$ which is multiplicative on disjoint unions and that can be defined using the rank-nullity formula on connected graphs, 
\begin{equation}
P_{\mbox{\tiny Tutte}}(x,y,\gamma)=\sum_{A\subset E}(y-1)^{n(A)}(x-1)^{r(E)-r(A)},
\end{equation}
where the sum runs over all subsets of the set of edges $E$ of $\gamma$. In the QFT language, the nullity  and the rank of a connected diagram can be expressed in terms of the number of internal lines and loops
\begin{equation}
\left\{
\begin{array}{rcl}
n(\gamma)&=&L_{\gamma}\cr
r(\gamma)&=&I_{\gamma}-L_{\gamma}
\end{array}
\label{ranknullity}
\right.
\end{equation}
Moreover, subsets of the sets of edges are in one-to-one correspondence with possibly disconnected subdiagrams, so that the Tutte polynomial can be written as the evaluation at $s=1$ of the character
\begin{equation}
\alpha=\exp_{\ast}s\left\{\delta_{\mathrm{tree}}+(y-1)\,\delta_{\mathrm{loop}}\right\}\ast
\exp_{\ast}s\left\{(x-1)\,\delta_{\mathrm{tree}}+\delta_{\mathrm{loop}}\right\},
\label{tutteprod}
\end{equation}
where we have used \eqref{ranknullity} and \eqref{diagexpo} to write the rank and nullity using the exponentials. From this expression, it is obvious that the Tutte polynomial character can be expressed as
\begin{equation}
\begin{array}{lcc}
\alpha&=&\overbrace{\exp_{\ast}s\left\{\delta_{\mathrm{tree}}+(y-1)\,\delta_{\mathrm{loop}}\right\}\ast
\exp_{\ast}s\left\{-\delta_{\mathrm{tree}}+\delta_{\mathrm{loop}}\right\}}^{\mbox{\tiny Tutte polynomial character at {\it x}=0 on the subgraphs}}\cr
&&\ast\cr
&&
\underbrace{\exp_{\ast}s\left\{\delta_{\mathrm{tree}}-\delta_{\mathrm{loop}}\right\}\ast
\exp_{\ast}s\left\{(x-1)\,\delta_{\mathrm{tree}}+\delta_{\mathrm{loop}}\right\}}_{\mbox{\tiny Tutte polynomial character at {\it y}=0 on the reduced graphs}},
\end{array}
\end{equation}
which is nothing but the convolution identity of Kook, Reiner and Stanton (see \cite{KRS}).

Then, using \eqref{diffFeyn}, $\alpha$ can be characterized as the solution of the differential equation
\begin{equation}
\frac{d\alpha}{ds}
=x\,\alpha\ast\delta_{\mathrm{tree}}
+y\,\delta_{\mathrm{loop}}\ast\alpha
+\left[\delta_{\mathrm{tree}},\alpha\right]_{\ast}
-\left[\delta_{\mathrm{loop}},\alpha\right]_{\ast}
\label{difftutte}
\end{equation}
that reduces to the trivial character at $s=0$. 

This characterisation of the Tutte polynomial as the solution of a differential equation can be used to derive its universality property. The latter is formulated  as follows. Consider a four variables graph polynomial $Q(x,y,a,b,\gamma)$ that is multiplicative on disjoint unions and one vertex unions (i.e. diagrams joined at a single vertex), and has the following  behavior under contraction and deletion of an arbitrary internal line:
\begin{itemize}
\item
if $\gamma'$ is obtained by deleting in $\gamma$ a bridge, then
\begin{equation}
Q(x,y,a,b,\gamma)=x\,Q(x,y,a,b,\gamma'),
\end{equation}
\item
if $\gamma'$ is obtained by contracting in $\gamma$ a self-loop, then
\begin{equation}
Q(x,y,a,b,\gamma)=y\,Q(x,y,a,b,\gamma'),
\end{equation}
\item
if $\gamma'$ is obtained by deleting in $\gamma$ a regular line (i.e. a line that is neither a bridge nor a self-loop) and $\gamma''$ obtained by contracting the same line, then
\begin{equation}
Q(x,y,a,b,\gamma)=a\,Q(x,y,a,b,\gamma')+b\,Q(x,y,a,b,\gamma'').
\end{equation}
\end{itemize}
Then, the character defined by
\begin{equation} 
\beta(\gamma)=s^{I_{\gamma}}\,Q(x,y,a,b,\gamma)
\end{equation}
obeys the differential equation
\begin{equation}
\frac{d\beta}{ds}
=x\,\beta\ast\delta_{\mathrm{tree}}
+y\,\delta_{\mathrm{loop}}\ast\beta+a\,\left[\delta_{\mathrm{tree}},\beta\right]_{\ast}
-b\,\left[\delta_{\mathrm{loop}},\beta\right]_{\ast}.
\end{equation}
This follows from the discussion at the end of last section, where we  have formulated algebraically the contraction/deletion of internal lines. Note that the contraction of a self-loop cancels in the fourth term  while the deletion of a bridge cancel in the third term because of the multiplicativity on one vertex unions. Acting with the automorphism $\varphi_{a^{-1}\!\!,b^{-1}}$ defined in \eqref{auto}, we obtain the Tutte polynomial differential equation \eqref{difftutte} with modified parameters $\frac{x}{a}$ and $\frac{y}{b}$
\begin{eqnarray}
\renewcommand{\baselinestretch}{4}
\frac{d}{ds}\varphi_{a^{-1}\!\!,b^{-1}}(\beta)
&=&\frac{x}{a}\,\varphi_{a^{-1}\!\!,b^{-1}}(\beta)\ast\delta_{\mathrm{tree}}
+\frac{y}{b}\,\delta_{\mathrm{loop}}\ast\varphi_{a^{-1}\!\!,b^{-1}}(\beta)\cr
&&+\,\left[\delta_{\mathrm{tree}},\varphi_{a^{-1}\!\!,b^{-1}}(\beta)\right]_{\ast}
-\,\left[\delta_{\mathrm{loop}},\varphi_{a^{-1}\!\!,b^{-1}}(\beta)\right]_{\ast}.
\end{eqnarray}
Accordingly, $\beta=\varphi_{a,b}(\alpha)$, with $\alpha$ the corresponding Tutte polynomial character, so that
\begin{equation}
Q(x,y,a,b,\gamma)=a^{r(\gamma)}b^{n(\gamma)}P_{\mbox{\tiny Tutte}}(\frac{x}{a},\frac{y}{b},\gamma).
\end{equation}
This is the universality property of the Tutte polynomial: Any graph polynomial multiplicative on disjoint and one vertex unions satisfying some simple deletion/contraction identities can be expressed using the Tutte polynomial.

As a second illustration, let us derive the relation of the Tutte polynomial with the $q$-state Potts model partition function using Feynman diagrams and effective actions. To this aim, we write the Tutte polynomial as a product of two characters \eqref{tutteprod} that we interpret as a composition of two background field effective action computation using \eqref{composeeff}. To simplify the notations, we set $u=x-1$, $v=y-1$ and $q=uv$, so that the factorization of the Tutte polynomial character reads, at $s=1$,
\begin{equation}
\alpha=\exp_{\ast}\left\{\delta_{\mathrm{tree}}+v\,\delta_{\mathrm{loop}}\right\}\ast
\exp_{\ast}\left\{u\,\delta_{\mathrm{tree}}+\delta_{\mathrm{loop}}\right\}.
\end{equation}
 Besides, for our purpose it is only necessary to consider  effective actions for a single 0 dimensional field which is nothing but a real variable. The first character is  $\exp_{\ast}\left\{\delta_{\mathrm{tree}}+v\,\delta_{\mathrm{loop}}\right\}$ and its induces a transformation of effective actions $S\mapsto S'$ that weights the Feynman diagrams with the character $\alpha(\gamma)=v^{L_{\gamma}}$. Then, using \eqref{loopeff} we get
\begin{equation}
S'[\phi]=v\log\left\{
\int[D\chi]\,\mathrm{e}^{-\frac{1}{2v}\chi^{2}}\,
\,\mathrm{e}^{\frac{S[\phi+\chi]}{v}}
\right\}.
\end{equation} 
Next, we consider $S'$ as a starting point for a new effective action computation $S'\rightarrow S''$ with a weight $\exp_{\ast}\left\{u\,\delta_{\mathrm{tree}}+\delta_{\mathrm{loop}}\right\}$, which amounts to weight internal lines by $u$ and loops by $\frac{1}{u}$, so that
\begin{equation}
S''[\psi]=\frac{1}{u}\log\left\{
\int[D\phi]\,\mathrm{e}^{-\frac{1}{2}\phi^{2}}\,
\,\mathrm{e}^{uS[\psi+\phi]}
\right\}.
\end{equation} 
As we replace $S'$ by its expression in terms of $S$ in the last equation, we get
\begin{equation}
S''[\psi]=\frac{1}{u}\log\left\{
\int[D\phi]\,\mathrm{e}^{-\frac{1}{2}\phi^{2}}\,
\left(
\int[D\chi]\,\mathrm{e}^{-\frac{1}{2v}\chi^{2}}\,
\,\mathrm{e}^{\frac{S[\psi+\phi+\chi]}{v}}
\right)^{q}
\right\}.
\end{equation} 
According to the general rule of effective action composition \eqref{composeeff}, this amounts to a single effective action computation with Feynman diagrams weighted by their Tutte polynomial. In particular, if we take a universal action like $S[\phi]=\lambda\mathrm{e}^{\phi}$ and evaluate $S''$ at $\psi=0$, we get the Tutte polynomial multiplied by a power of $\lambda$
\begin{equation}
\frac{1}{u}\log\left\{
\int[D\phi]\,\mathrm{e}^{-\frac{1}{2}\phi^{2}}\,
\left(
\int[D\chi]\,\mathrm{e}^{-\frac{1}{2v}\chi^{2}}\,
\mathrm{e}^{\frac{\lambda}{v}\mathrm{e}^{\phi+\chi}}
\right)^{q}
\right\}=\sum_{\gamma\,\mathrm{connected}}
\frac{\lambda^{V_{\gamma}}}{\mathrm{S}_{\gamma}}P_{\mbox{\tiny Tutte}}(u+1,v+1,\gamma),
\label{generating}
\end{equation} 
with $V_{\gamma}$ the number of vertices of $\gamma$.
The choice of the universal action $S[\phi]=\lambda\mathrm{e}^{\phi}$ is motivated by the fact that it generates all diagrams with a factor 1. If restrict ourselves to a monomial like $\lambda\frac{\phi^{n}}{n!}$, then we generate only the diagrams with $n$-valent vertices. 

Unfortunately, it is not yet possible to interpret \eqref{generating} as a generating function for the Tutte polynomial, since this would require us to be able to identify the contribution of each diagram in the  perturbative expansion. To proceed, we have to write the integrals over $\phi$ and $\chi$ as a single integral which is a perturbation of a Gau\ss ian integral.   This can be easily done when $q$ is an integer by introducing $q$ independent fields $\chi$ 
\begin{equation}
\left(
\int[D\chi]\,\mathrm{e}^{-\frac{1}{2v}\chi^{2}}\,
\,\mathrm{e}^{\frac{1}{v}S[\phi+\chi]}
\right)^{q}
=\int \prod_{1\leq i\leq q}[D\chi_{i}]\,
\mathrm{e}^{-\frac{1}{2v}\sum_{i}(\chi_{i})^{2}}\,\mathrm{e}^{\frac{1}{v}\sum_{i}S[\chi_{i}+\phi]}
\end{equation}
Then, the integral over $\phi$ and $\chi$ in \eqref{generating} reads
\begin{equation}
\int[D\phi]\int \prod_{1\leq i\leq q}[D\chi_{i}]\,
\mathrm{e}^{-\frac{1}{2}\phi^{2}}
\mathrm{e}^{-\frac{1}{2v}\sum_{i}(\chi_{i})^{2}}\mathrm{e}^{\frac{1}{v}\sum_{i}S[\chi_{i}+\phi]}
\end{equation}
It is convenient to trade $\chi_{i}$ for $\xi_{i}=\chi_{i}+\phi$ so that the integral over $\phi$ is Gau\ss ian 
\begin{equation}
\int [D\phi]\int \prod_{1\leq i\leq q}[D\xi_{i}]\,
\mathrm{e}^{-\frac{1}{2}\left(1+u\right)\phi^{2}
+\frac{\phi}{v}\sum_{i}\xi_{i}}
\times\mathrm{e}^{-\frac{1}{2v}\sum_{i}\xi_{i}^{2}}
\times\mathrm{e}^{\frac{1}{v}\sum_{i}S[\xi_{i}]}.
 \end{equation}
Performing the Gau\ss ian integral over $\phi$  and discarding an overall constant which is absorbed in the normalization, we get
\begin{equation}
\int \prod_{1\leq i\leq q}[D\xi_{i}]\,
\mathrm{e}^{-\frac{1}{2v}\left\{
\left(\sum_{i}(\xi_{i})^{2}-\frac{1}{v(1+u)}\left(\sum_{i}\xi_{i}\right)^{2}\right)\right\}}
\times\mathrm{e}^{\frac{1}{v}\sum_{i}S[\xi_{i}]}.
 \end{equation}
This can be written as a perturbed Gau\ss ian integral over a multiplet of fields $\xi=(\xi_{i})$, so that \eqref{generating} reads
\begin{equation}
\frac{1}{u}\log\left\{
\int  [D\xi]\,
\mathrm{e}^{
-\frac{1}{2}\xi\cdot A^{-1}\xi}\, \mathrm{e}^{V(\xi)}
\right\}=
\sum_{\gamma\,\mathrm{connected}}
\frac{\lambda^{V_{\gamma}}}{\mathrm{S}_{\gamma}}P_{\mbox{\tiny Tutte}}(u+1,v+1,\gamma),
\label{generating2}
\end{equation}
with a $q\times q$ propagator
\begin{equation}
A=v+M
\end{equation}
where $M$ is the $q\times q$ matrix whose entries are all equal to 1 and an interaction
\begin{equation}
V(\xi)=\frac{\lambda}{v}\sum_{i}\exp(\xi_{i}).
\end{equation}
The expansion of \eqref{generating2} over Feynman diagrams can be considered as a generating functions for the Tutte polynomials, since each diagram is weighted by $P_{\mbox{\tiny Tutte}}(u+1,v+1,\gamma)$. Moreover, the Feynman rules for this expansion are as follows,  
\begin{itemize}
\item
each vertex contributes to a factor $\frac{\lambda}{v}$ and is equipped with an index $i$ taking $q$ values, because of the $q$ independent fields in the interaction; 
\item
each propagator contributes a factor of $1+v$ if it connects vertices with the same index and a factor of 1 otherwise,
\item
sum over all the indices. 
\end{itemize}
Up to a factor, this is  nothing but the partition function $Z(\beta,J,\gamma)$ for the $q$-state Potts model defined on the graph $\gamma$. Recall that the latter is a lattice model that is defined on an arbitrary graph $\gamma$ by assigning spins $\sigma_{v}$ taking $q$ values to the vertices of the diagram. The energy of a configuration of spins $\sigma=(\sigma_{v})$ is defined as a sum over all the edges of the diagram 
\begin{equation}
H(\sigma)=-J\sum_{\mathrm{edges}\,e}\delta_{\sigma_{v},\sigma_{v}'},
\end{equation} 
with $v$ and $v'$ the vertices linked by $e$ and $J>0$ a constant that favors spin alignment. Its partition function is defined as
\begin{equation}
Z(\beta,J,\gamma)=\sum_{\sigma}\mathrm{e}^{-\beta H(\sigma)}
\end{equation}
and it follows from the previous discussion that it is proprtional to the Tutte polynomial 
\begin{equation}
Z(\beta,J,\gamma)=uv^{V_\gamma}\,P_{\mbox{\tiny Tutte}}(u+1,v+1,\gamma)
\end{equation}
with $v=\mathrm{e}^{-\beta J}-1$.

\subsection{Iterative solution of Polchinski's equation}
\label{polchsec}

Let us now come back to Polchinski's equation \eqref{erge} that describes the scale evolution of the effective action $S_{\Lambda}$. Recall that $S_{\Lambda}$ is obtained from a bare action $S_{\Lambda_{0}}$ valid at a very high energy scale by integrating Fourier modes between $\Lambda$ and $\Lambda_{0}$. A large class of propagators that implement this integration and preserve both translational and rotational symmetry can be constructed in momentum space as
\begin{equation}    
A_{\Lambda,\Lambda_{0}}(p.q)=
\left\{\chi\left(\frac{p^{2}}{\Lambda_{0}^{2}}\right)-\chi\left(\frac{p^{2}}{\Lambda^{2}}\right)\right\}
\times\frac{\delta(p+q)}{p^{2}}.
\end{equation}
The cut-off function $\chi$ a smooth positive decreasing function with values close to 1 on $[0,1]$ and close to 0 on $[1,+\infty[$. A typical example is $\chi(s)=\mathrm{e}^{-s}$ that yields a propagator 
\begin{equation}
A_{\Lambda,\Lambda_{0}}(p,q)=\delta(p+q)\int^{\frac{1}{\Lambda^{2}}}_{\frac{1}{(\Lambda_{0})^{2}}}\!\!d\alpha\,\,\mathrm{e}^{-\alpha p^{2}},
\end{equation}
that leads to the convenient expression of Feynman diagrams based on Symanzik's polynomials for $\phi^{4}$ theory (see \cite{Rivasseau} and the remark at the end of this section). Note that we have not included the mass term in the propagator to simplify the forthcoming dimensional analysis. Here we treat the mass term as an interaction, which does not modify the UV behavior of the theory.  The derivative is independent of $\Lambda_{0}$
\begin{equation}
\Lambda\frac{\partial A_{\Lambda,\Lambda_{0}}}{\partial\Lambda}(p,q)=
\frac{2}{\Lambda^{2}}\chi'\left(\frac{p^{2}}{\Lambda^{2}}\right)
\times\delta(p+q).
\end{equation}
and implements the integration over an infinitesimal shell.  With such a propagator, Polchinski's equation can be cast in the general form \eqref{generalRGE} with the non linear operator
\begin{equation}
\beta(\Lambda,S)=\int dpdq\,\Lambda\frac{\partial A_{\Lambda,\Lambda_{0}}}{\partial\Lambda}(p,q)\,
\left(
\frac{\delta^{2}S}{\delta\widetilde{\phi}(p)\delta\widetilde{\phi}(q)}- \frac{\delta
S}{\delta\widetilde{\phi}(p)} \frac{\delta S}{\delta\widetilde{\phi}(q)} \right)
\end{equation}

In high energy physics with $c=\hbar=1$, all quantities are measured in terms of masses so that if we scale all masses by a factor $\mathrm{e}^{s}$, the action transforms as $S\rightarrow S'=\mathrm{e}^{s{\cal D}} S$ with $S'[\mathrm{e}^{s(D/2+1)}\widetilde{\phi}']=S[\widetilde{\phi}]$ and $\widetilde{\phi}'(p)=\widetilde{\phi}(\mathrm{e}^{s}p)$ in a space-time of dimension $D$. Then, it is easy to check that 
\begin{equation}
\beta(\mathrm{e}^{s}\Lambda,\mathrm{e}^{s{\cal D}} S)=\mathrm{e}^{s{\cal D}}\beta(\Lambda,S),
\end{equation}
 so that all the results of section \ref{diffren} apply. In particular, a translation invariant term of the form
\begin{equation}
\int dp_{1}\cdots dp_{N}\,\delta(p_{1}+\cdots+p_{N})\,\Gamma(p_{1},\dots,p_{N})\,
\widetilde{\phi}(p_{1})\cdots\widetilde{\phi}(p_{N})
\end{equation}
where $\Gamma$ is homogenous of degree $n$, scales with a factor $\mathrm{e}^{sd }$ with
\begin{equation}
d=D-N\left(\frac{D}{2}-1\right)-n.
\end{equation}
Therefore, we recover the fact that the mass term is always relevant and that for $\phi^{4}$ theory in dimension 4 the coupling constant is marginal and all terms with a high number of fields and/or high order derivative interactions are irrelevant.

Accordingly, Polchinski's equation can be written in integral form,
\begin{equation}
S_{\Lambda}=S_{\Lambda_{0}}+\int_{\Lambda_{0}}^{\Lambda}d\Lambda'
\beta(\Lambda',S_{\Lambda'})
\end{equation}
with $\dot{A}=\frac{\partial A}{\partial \Lambda}$ and
\begin{equation}
\beta(\Lambda,S)
=\frac{1}{2}
\int\!dpdq\,\,  \dot{A}_{\Lambda}(p,q)
\left(
\frac{\delta^{2}S}{\delta\widetilde{\phi}(p)\delta\widetilde{\phi}(q)}
- \frac{\delta S}{\delta\widetilde{\phi}(p)} \frac{\delta S}{\delta\widetilde{\phi}(q)} \right)
\end{equation}
Then, its iterative solution is expanded over ordered Feynman diagrams. The latter are Feynman diagrams with a hierarchy of internal lines defined by drawing boxes on the diagram such that the boxes are either disjoint or nested and such that a box differ from the next one it is included in by an internal line. Alternatively, the order on the internal lines can be pictured by a rooted tree $t$ drawn on the set of internal lines of the diagram. The contribution $A_{\Lambda,\Lambda_{0}}^{\gamma_{\mathrm{ord}}}(S_{0})$ to the effective action $S_{\Lambda}$ is computed as follows,
\begin{itemize}
\item
associate an intermediate cut-off $\Lambda_{i}$ to any internal line;
\item
compute the value of the Feynman diagram using the background field Feynman rules with vertices derived from $S_{0}$ and propagators $\dot{A}_{\Lambda_{i}}$;
\item
integrate the variables $\Lambda_{i}$ over the treeplex $I_{\Lambda,\Lambda_{0}}^{t}$, defined in section \ref{difftime}.
\end{itemize}
Accordingly, the low energy effective action reads
\begin{equation}
S_{\Lambda}=\sum_{\gamma_{\mathrm{ord}}}\frac{A_{\Lambda,\Lambda_{0}}^{\gamma_{\mathrm{ord}}}(S_{0})}{\mathrm{S}_{\gamma_{\mathrm{ord}}}}
\end{equation}
with $\mathrm{S}_{\gamma_{\mathrm{ord}}}$ the cardinal of the automorphisms  group of the diagram that preserve the ordering.

Ordered diagrams generate a commutative Hopf algebra
${\cal H}_{F}^{\mathrm{ord}}$ with coproduct similar to
that of ${\cal H}_{F}$ given in \eqref{coproductfeyn}, except that it has to preserve the box structure. This coproduct is very similar to the rooted tree coproduct, as it reproduces on the diagram the admissible cuts of the tree as an extraction of disjoint boxes. For example, with $\Delta'$ the non trivial part of
$\Delta$:
\begin{center}
\includegraphics[width=12cm]{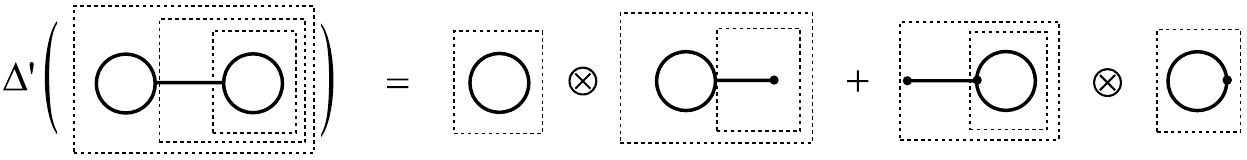},\\
\includegraphics[width=6cm]{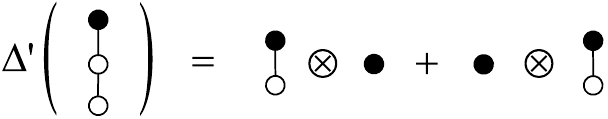},\\
\includegraphics[width=12cm]{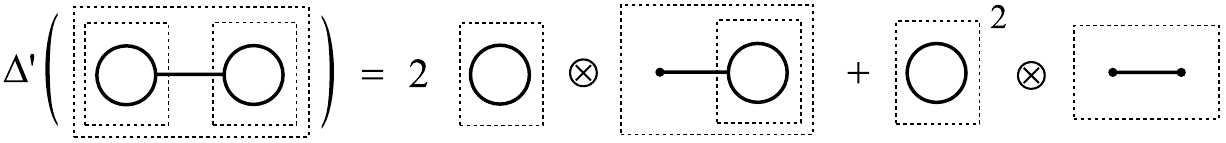},\\
\includegraphics[width=6cm]{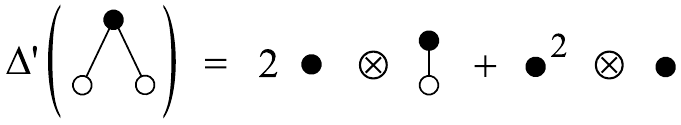}.\\
\end{center}
Because the effective action could as well be computed in a single step using the propagator $A_{\Lambda,\Lambda_{0}}$, 
\begin{equation}
\sum_{\gamma}\frac{A_{\Lambda,\Lambda_{0}}^{\gamma}(S_{0})}{\mathrm{S}_{\gamma}}
=\sum_{\gamma_{\mathrm{ord}}}\frac{A_{\Lambda,\Lambda_{0}}^{\gamma_{\mathrm{ord}}}(S_{0})}{\mathrm{S}_{\gamma_{\mathrm{ord}}}},
\end{equation}
there is a Hopf algebra morphism from ${\cal H}_{F}$ to ${\cal H}_{F}^{\mathrm{ord}}$ defined by summing over all the orders on the diagram,
\begin{equation}
\frac{\gamma}{{\mathrm{S}}_{\gamma}}\rightarrow
\mathop{\sum}\limits_{\mathrm{ orders\, on}\,\gamma}
\frac{\gamma^{\mathrm{ord}}}{{\mathrm{
S}}_{\gamma}^{^{\mathrm{ord}}}}.
\end{equation}
Moreover, there is also a morphism from the algebra ${\cal H}_{F}^{\mathrm{ord}}$ to the algebra of rooted trees which associates n ordered diagram with its tree
\begin{equation}
\gamma^{\mathrm{ord}}\rightarrow t
\end{equation}
If we pullback  the tree factorial by the composition of these two morphisms we get
\begin{equation}
\frac{1}{{\mathrm{S}}_{\gamma}}\rightarrow
\mathop{\sum}\limits_{\mathrm{ orders\, on}\,\gamma}
\frac{1}{t!\,{\mathrm{
S}}_{\gamma}^{^{\mathrm{ord}}}},
\end{equation}
in accordance with the combinatorial interpretation of the tree factorial in \eqref{combifac}.

\begin{figure}
\[
\begin{array}{ccc}
\includegraphics[width=5cm]{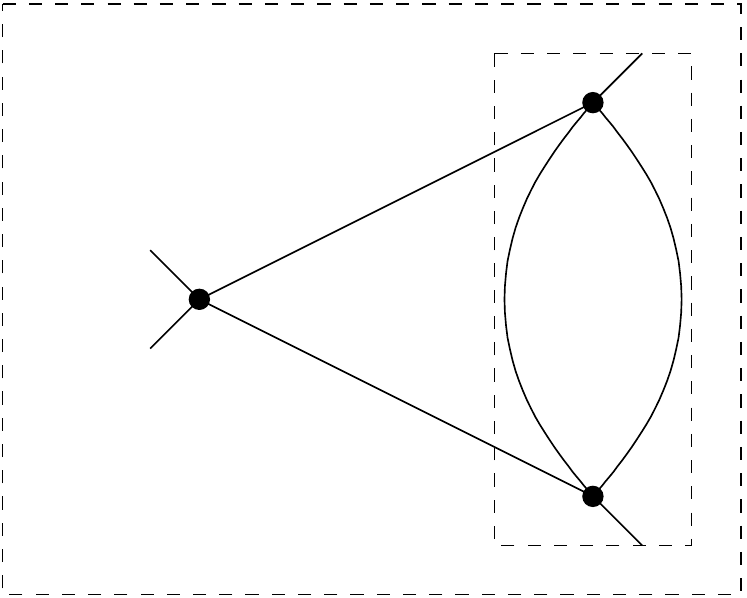}
&\quad&
\includegraphics[width=5cm]{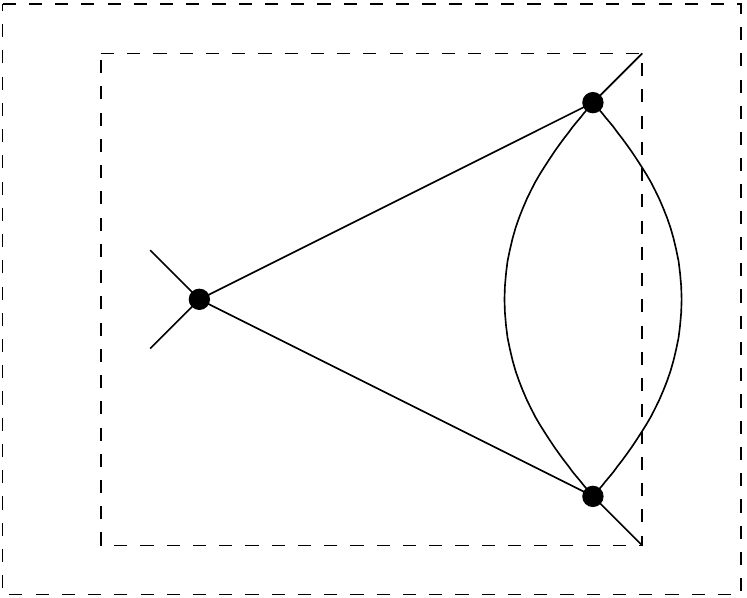}\cr
\propto(\log\Lambda)^{2}&\quad&\propto\log\Lambda
\end{array}
\]
\caption{Two decompositions of the one loop four point function}
\label{fourpoint}
\end{figure}

From a renormalization group viewpoint, the use of ordered diagrams is necessary since a given diagram can be decomposed as different ordered ones that behave slighty differently.  For instance, the 2 loop contribution to effective low energy coupling in $\phi^{4}$ theory in 4 dimensions depicted in figure \ref{fourpoint} can be ordered in two kind of ways: Either we consider it a 4 point function inserted in a 4 point function or as a 6 point function inserted in a 2 point function. In the first case we get a leading contribution proportional to $(\log\Lambda)^{2}$ whereas in the second case it is proportional $\log\Lambda$. 

To implement dimensional analysis at the level of the Hopf algebras  ${\cal H}_{F}$ and ${\cal H}_{F}^{\mathrm{ord}}$ we introduce decorated diagrams with generalized external structures. Let us decompose the space of action functionals as a direct sum ${\cal E}={\cal E}'\oplus{\cal E}''$, with ${\cal E}'$ containing the relevant and marginal terms, which span a finite dimensional space. The decorations are labels $i\in I$ for a basis of eigenstates of the scaling operator ${\cal D}$ in ${\cal E}'$, which we write as ${\cal O}_{i}$ and an extra index for ${\cal E}''$. For instance, for $\phi^{4}$ theory in dimension 4, a basis of ${\cal E}'$ is given by the operators
\begin{equation}
\left\{\int d^{4}x\, \frac{\phi^{4}}{4!},\,\int d^{4}x \frac{\phi^{2}}{2},\int d^{4}x\, \frac{(\partial\phi)^{2}}{2}\right\}
\end{equation}
and we can identify the coupling constant, square mass and wave function parameter as the components of the action in this basis. These labels are generalization of the external structures introduced by Connes and Kreimer in \cite{ck1}, the main difference being that we have to treat all the interactions and not only the renormalizable ones. 

A decorated Feynman diagram with a generalized external structure (with an order on the internal lines or not) is a couple $(\widetilde{\gamma},i)$ where the vertices of $\widetilde{\gamma}$ are decorated with indices in $I$. These diagrams generate a commutative Hopf algebra ${\cal H}_{F}^{\mathrm{dec}}$ with coproduct  
\begin{equation}
\Delta(\widetilde{\gamma},i)=(\widetilde{\gamma},i)\otimes 1+1\otimes(\widetilde{\gamma},i)+
\mathop{\sum}\limits_{\widetilde{\gamma}_{1},\dots,\widetilde{\gamma}_{n}
{\mathrm{disjoint}\,\subset\widetilde{\gamma}}
\atop
i_{1},\dots,i_{n} \,\mathrm{decorations}\,\in I}
(\widetilde{\gamma}_{1},i_{1})\cdots(\widetilde{\gamma}_{n},i_{n})
\otimes\left(\frac{\widetilde{\gamma}}{(\widetilde{\gamma}_{1},i_{1})\cdots(\widetilde{\gamma}_{n},i_{n})},i\right),
\end{equation}
with the vertices of the reduced diagram decorated by the generalized external structures of the subdiagrams. 

Using the background field technique, such a diagram is evaluated with vertices obtained by projecting onto the eigenspace of the decoration carried by the vertex for the fluctuating field $\chi$ and the final result projected onto the eigenspace associated to $i$ for the background field $\phi$. Then, all the composition identities such as \eqref{composeeff} hold since they simply amount to inserting the decomposition ${\cal E}={\cal E'}\oplus{\cal E}$ for each action functional. 

The unrenormalized low energy effective action $S_{\Lambda}^{\mathrm{un}}$ is computed by integrating modes with a propagator $A_{\Lambda,\Lambda_{0}}$ starting with a bare action $S_{\Lambda_{0}}$, so that
\begin{equation} 
S_{\Lambda}^{\mathrm{un}}=\Psi_{\alpha}^{A_{\Lambda,\Lambda_{0}}}(S_{\Lambda_{0}}),\label{un}
\end{equation}  
where $\alpha$ is the character that takes the value 1 on all diagrams. The renormalization procedure amounts to choose $S_{\Lambda_{0}}$ in such a way that at the low energy scale $\Lambda$ the relevant and marginal parameters are fixed to $S_{\mathrm{R}}$. If we denote by $\beta$ the character that takes the value 1 on the diagram with relevant and marginal external structures, then the bare action is determined by
\begin{equation} 
S_{\mathrm{R}}=\Psi_{\beta}^{A_{\Lambda,\Lambda_{0}}}(S_{\Lambda_{0}}),
\end{equation}  
Then, inverting this equation and substituting in \eqref{un}, we get the renormalized low energy effective theory as
\begin{equation} 
S_{\Lambda}^{\mathrm{ren}}=\Psi_{\beta^{-1}\ast\alpha}^{A_{\Lambda,\Lambda_{0}}}(S_{\mathrm{R}}).
\end{equation}  
In this discussion, we have used the characters only as a means of selecting some of the diagrams. In fact, the evaluation of the diagrams themselves leads to characters thanks to the decorations and external structures. This is the point of view adopted by Connes and Kreimer and is definitely more powerful since it allows to use the algebraic structure based on the Hopf algebra. Unfortunately, in our case it leads to a rather complicated formalism when we compose two effective action computations since we have to keep track of the location of the insertion of diagrams. 

Finally, let us conclude by a simple example of an application of the previously outlined Lie algebraic techniques to the computation of the first Symanzik polynomial. In quantum field theory, a Feynman diagram $\gamma$ with $n$ edges can be evaluated, in dimension $D$, as
\begin{equation}
\int \frac{d^{n}\alpha}{\big[\textstyle{U_{\gamma}(\alpha)}\big]^{\frac{D}{2}}}\,\mathrm{e}^{-\frac{V_{\gamma}(\alpha,p)}{U_{\gamma}(\alpha)}}
\end{equation}
where $U_{\gamma}(\alpha)$ and $V_{\gamma}(\alpha,p)$ are polynomials in the Schwinger parameters $\alpha=(\alpha_{1},\dots,\alpha_{n})\in{\Bbb R}^{n}$ and the momenta $p$. Explicitly, the first Symanzik polynomial is expressed as 
\begin{equation}
U_{\gamma}(\alpha)=\sum_{t\atop \mbox{\tiny spanning trees}}\prod_{i\notin t}\alpha_{i},
\end{equation}
where a spanning tree is a tree drawn on $\gamma$ that touches all the vertices.

\begin{figure}
\begin{center}
\includegraphics[width=5cm]{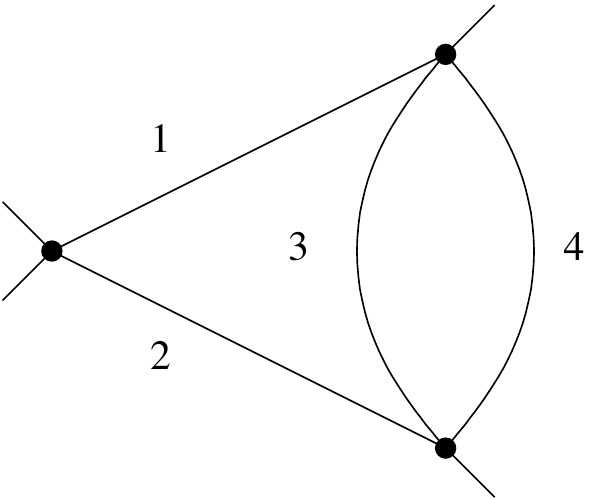}
\caption{A two loop diagram with labelled edges}
\label{fourpointlab}
\end{center}
\end{figure}
For example, for the diagram depicted in figure \ref{fourpointlab}, the first Symanzik polynomial is 
\begin{equation}
U_{\includegraphics[width=0.4cm]{fourpointnum.eps}}(\alpha)=
\alpha_{1}\alpha_{3}+\alpha_{1}\alpha_{4}+\alpha_{2}\alpha_{3}+\alpha_{2}\alpha_{4}+\alpha_{3}\alpha_{4}.
\end{equation}

In the Hopf algebra of Feynman diagrams with labelled edges consider the infinitesimal characters such that
\begin{equation}
\delta_{\mathrm{tree}}(\parbox{0.4cm}{\mbox{\includegraphics[width=0.4cm]{tutte8.eps}}})=1
\qquad\mathrm{and}\qquad
\delta_{\mathrm{loop}}(\parbox{0.4cm}{\mbox{\includegraphics[width=0.4cm]{tutte5.eps}}})=\alpha_{i}
\end{equation}
and and that vanish otherwise.  Then, the first Symanzik polynomial can be written as
\begin{equation}
U_{\gamma}(\alpha)=
\mathrm{e}^{\delta_{\mathrm{tree}}}\ast
\mathrm{e}^{\delta_{\mathrm{loop}}}
\end{equation}
Indeed, the first factor extracts all the trees and contracts them. Then, the second factor ensures that the reduced diagram has a unique vertex, which means that we extract only spanning trees, and assigns to it the product of the $\alpha$ over the non contracted edges. Then, 
\begin{equation}
U_{\gamma}(\alpha)=\mathrm{e}^{\delta_{\mathrm{tree}}}\ast\mathrm{e}^{\delta_{\mathrm{loop}}}\ast
\mathrm{e}^{-\delta_{\mathrm{tree}}}\ast\mathrm{e}^{\delta_{\mathrm{tree}}}
=
\mathrm{e}^{\sum_{n}\delta_{n\,\mathrm{loop}}}\ast\mathrm{e}^{\delta_{\mathrm{tree}}}
\end{equation}
where $\delta_{n\,\mathrm{loop}}$ is the infinitesimal character that takes the value $\sum_{i}\alpha_{i}$  
on the one loop diagram with $n$ edges and vanishes otherwise. This last relation follows from the Lie algebraic identitity,  
\begin{equation}
\frac{1}{n!}\underbrace{\Big[\delta_{\mathrm{tree}},\big[\cdots\left[\delta_{\mathrm{tree}},\delta_{\mathrm{loop}}\right]\cdots\big]\Big]}_{n\,\mathrm{iterations}}=\delta_{n\,\mathrm{loop}}
\end{equation}
interpreted in terms of the  contraction/deletion relations \eqref{diffrelations}.

Thus, $U_{\gamma}(\alpha)$ can be evaluated by summing over all contraction schemes of the loops, generated by the iteration of the coproduct in the computation of  $\mathrm{e}^{\sum_{n}\delta_{n\,\mathrm{loop}}}$. For example, for the diagram depicted in figure \ref{fourpointlab},
\begin{equation}
U_{\includegraphics[width=0.4cm]{fourpointnum.eps}}(\alpha)=
\frac{1}{2}\big[
(\alpha_{1}+\alpha_{2}+\alpha_{3})\alpha_{4}+
(\alpha_{1}+\alpha_{2}+\alpha_{4})\alpha_{3}+
(\alpha_{1}+\alpha_{2})(\alpha_{3}+\alpha_{4})
\big]
\end{equation}
These terms correspond to the following contraction schemes
\begin{equation}
\big\{(123,4),(124,3),(34,12)\big\}.
\end{equation}
In general, such an expression for the first Symanzik polynomial may prove useful in the study of the UV divergences, since the latter are governed by the integrals over loops.

\subsection{An algebraic structure related to Schwinger-Dyson equations}

\begin{figure}
\centerline{\includegraphics[width=13cm]{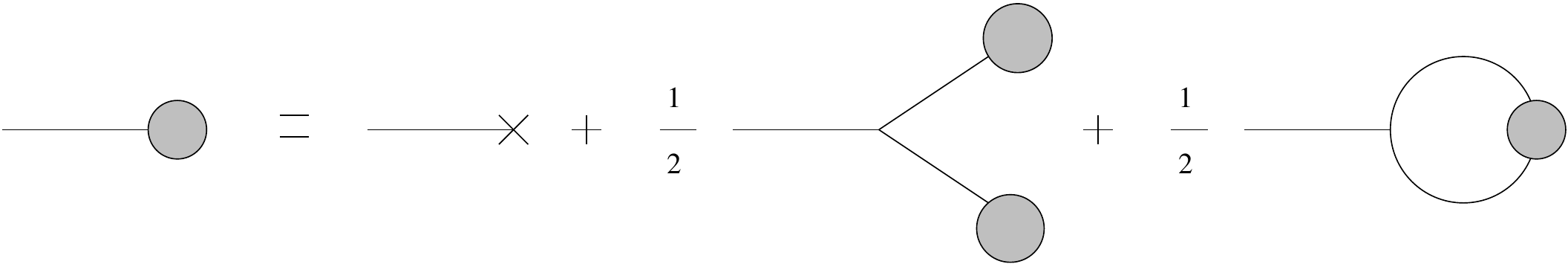}}
\caption{Schwinger-Dyson equation in $\phi^{3}$ theory}
\label{SDfig}
\end{figure}

Schwinger-Dyson equations are a set of functional differential equations playing a role analogous to the equation of motion in QFT (see \cite{cvitanovic} for a general discussion). When solved order by order, they allow to recover the perturbative expansion in terms of couplings constants. More precisely, they are functional differential equations for the derivative of the generating function ${\cal W}[J]$ of the connected Feynman diagrams.  ${\cal W}[J]$ is expressed as a functional of the source $J$ as
\begin{equation}
{\cal W}[J]=h\log{\cal Z}[J]
\end{equation}
with ${\cal Z}[J]$ defined as
\begin{equation}
{\cal Z}[J]={\cal N}\int[D\phi]\,\mathrm{e}^{\frac{-S[\phi]+J\cdot\phi}{h}}
\end{equation}
and the normalization constant defined such that ${\cal Z}[0]=0$. Note that we have introduced $h$ so that we can measure deviation from the classical theory obtained at $h=0$.

One of the structural properties of the path integral is its invariance under change of variables in the space of fields over which we integrate. In the simplest case of an infinitesimal translation $\phi(x)\rightarrow\phi(x)+\epsilon(x)$, the invariance of ${\cal Z}$ implies
\begin{equation}
\left(-S'\left[\frac{\delta}{\delta J}\right]+J\right){\cal Z}[J]=0
\end{equation} 
Replacing ${\cal Z}[J]$ by $\mathrm{e}^{\frac{{\cal W}}{h}}$, it yields an equation for $\frac{\delta{\cal W}}{\delta J}$ whose complexity increases with the powers of the monomials entering into $S$.
In the simplest case of the $\phi^{3}$ interaction,
\begin{equation}
S[\phi]=\int{d^{D}x}\,\left(\frac{1}{2}\phi\left(-\Delta+m^{2}\right)\phi+\frac{g}{3!}\phi^{3}\right),
\end{equation}
the Schwinger-Dyson equation reads
\begin{equation}
\frac{\delta{\cal W}}{\delta J(x)}=\int d^{D}y\left\{\,K(x,y)J(y)+
+\frac{g}{2}\,K(x,y)\left(\frac{\delta {\cal W}}{\delta J(y)}\right)^{2}
+\frac{gh}{2}\,K(x,y)\frac{\delta^{2}{\cal W}}{\delta J(y)\delta J(y)}\right\},\label{SD}
\end{equation}
with $K$ the kernel of the inverse of $-\Delta+m^{2}$. Let us represent the propagator  $K$ by a line, $J$ by a cross and ${\cal W}$ by a blob out of which we draw a line for each differenciation with respect to $J$. Then, eq. \eqref{SD} is neatly summarized in figure \ref{SDfig}. These equations generalize to more complicated interactions: if we consider a $\phi^{n}$ interaction we have a term for each partition of $n-1=n_{1}+\cdots+n_{k}$ corresponding to the product of the $n_{i}^{\mathrm{th}}$ functional derivatives of ${\cal W}$. Of course, the functional differentials are distributions and the equations may become singular when the latter are evaluated at coinciding points, leading again to the UV divergences. In this section, we shall ignore this problem, for instance by working with the regularized theory, and focus on the formal aspects of the Schwinger-Dyson equations. 
  
The equation \eqref{SD} is a fixed point equation for $\varphi=\frac{\delta {\cal W}}{\delta J}$ and may be solved using a geometric series over rooted  trees. Indeed, rewrite  \eqref{SD} as
\begin{equation}
\varphi=\varphi_{0}+{\cal F}_{\mathrm{tree}}\big[\varphi\big]+{\cal F}_{\mathrm{loop}}\big[\varphi\big].
\end{equation}
and expand $\varphi$ as a sum over rooted trees with black vertices for ${\cal F}_{\mathrm{tree}}$ and white ones for ${\cal F}_{\mathrm{loop}}$. Note that ${\cal F}_{\mathrm{tree}}$ is bilinear, so that black vertices have one or two outgoing edges whereas white ones have only one because ${\cal F}_{\mathrm{loop}}$ is linear. As we compose these operators along the trees, we obtain sums of Feynman diagrams with a distinguished external line and terminal lines equipped by crosses. If we iterate ${\cal F}_{\mathrm{tree}}$ alone we only obtain tree-level diagrams while ${\cal F}_{\mathrm{loop}}$ generates loops.

The correspondence between these Feynman diagrams and rooted trees expansions can be made more precise if we introduce a hierarchy on the vertices of the diagram, in complete analogy with the ordering of the lines in the iterative solution of Polchinski's equation.  The ordering of is encoded in a spanning rooted tree drawn on the diagram constructed as follows. The first vertex we meet starting from the distinguished external leg is defined to be the highest one and is the root of the tree.  As we remove this vertex from the diagram, the latter may fall into several connected components. If the 
diagrams is disconnected after the removal of the vertex, we color the root in black, otherwise we leave it blank. Next, in each component, we choose as distinguished external line any line that was previously connected to the vertex we removed. Then, the lines point towards new vertices that are just below the vertex we removed in the hierarchy. No order is assumed between the vertices belonging to different connected components. Then, we iterate the operation, considering any component as a new 
diagram with a distinguished vertex till we exhaust all vertices of the diagram. 

It is important to note that the trees we obtain have white vertices with at most one outgoing edge and  black vertices at most two. Besides, we add extra outgoing edges, called black leaves, to the black vertices so that they have exactly two outgoing edges. Then, the tree we obtain are such that below any white vertex, there is always less  white vertices than black leaves. This property is natural from the Schwinger-Dyson point of view:
The black leaves represent the sources $J$ and the white vertices the differential operator ${\cal F}_{\mathrm{loop}}$, which vanishes if there are not enough sources.  Conversely, given any tree with this property, we can reconstruct all the Feynman diagrams that appear in the corresponding power of the Schwinger-Dyson equation by relating each white vertex to a black leaf located below. This tree structure drawn on the Feynman diagrams may be used to define another Hopf algebra structure by performing admissible cuts on the tree. Of course, the correspondence between Feynman diagrams and trees is not one to one: To a tree correspond several diagrams and to a diagram several trees.

\begin{figure}\[
\parbox{7cm}{\includegraphics[width=7cm]{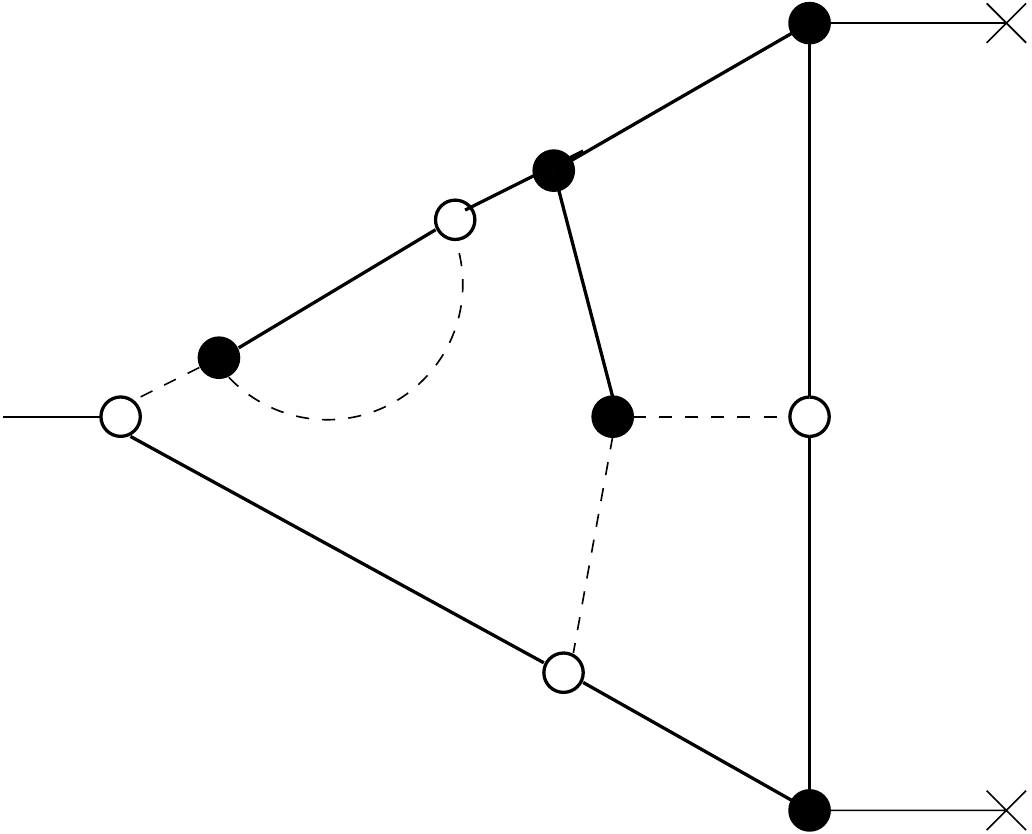}} \longleftrightarrow\qquad
\parbox{12cm}{\includegraphics[width=12cm]{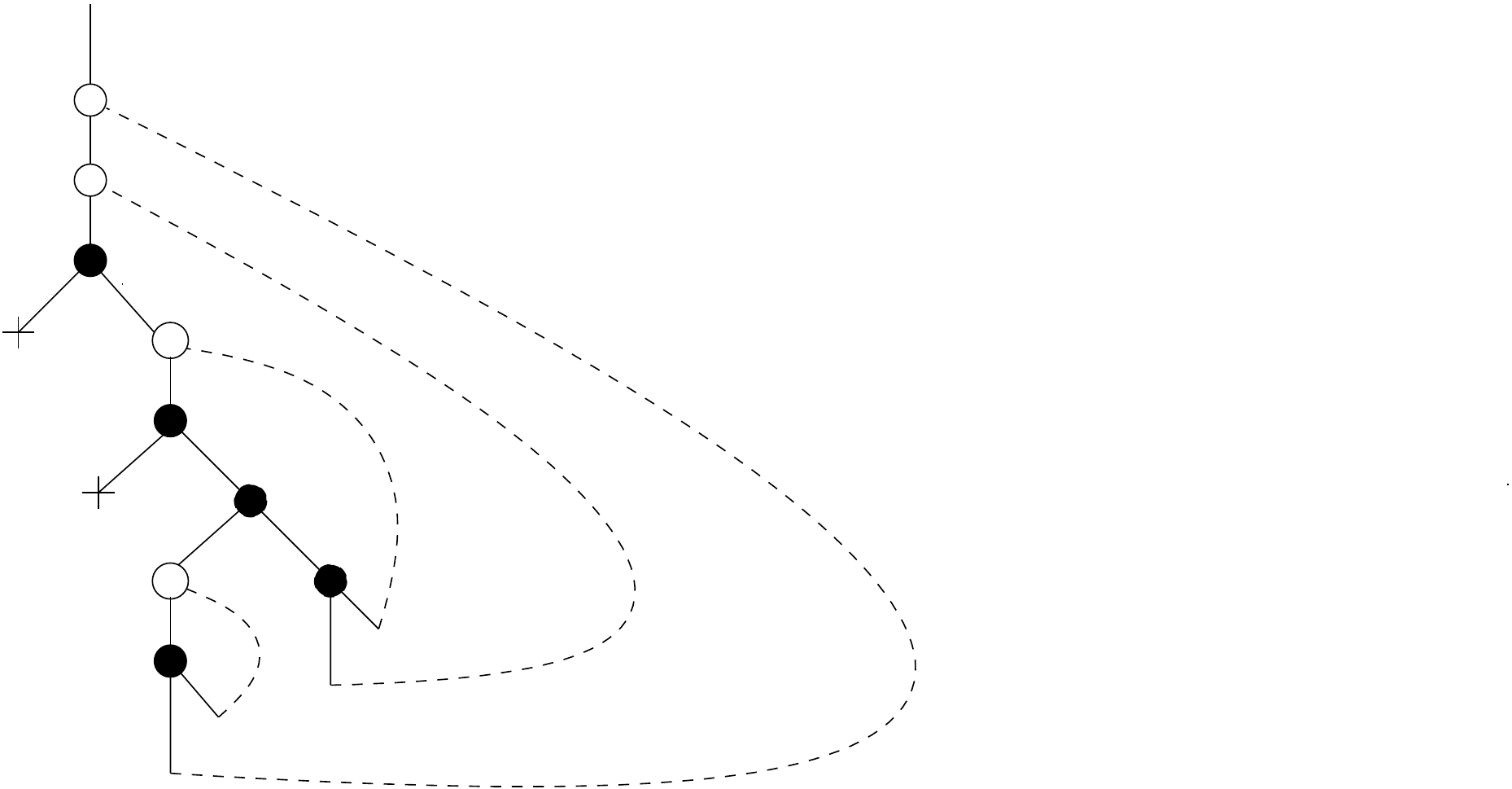}}\]
\caption{A planar $\phi^{3}$ diagram and its reconstruction from a tree}
\label{bijection}
\end{figure}

For planar diagrams, the ambiguity in the association of a rooted tree to a Feynman diagram can be disposed with by deciding to always move counterclockwise around the diagram as we choose the distinguished internal lines, starting from the distinguished external line. This way we associate a unique tree to each planar diagram. To reconstruct the Feynman diagram from the tree, we link every white vertex (starting with the lower ones) which is not a terminal one to the next black leaf above in the hierarchy of vertices, with a link always moving clockwise around the diagram.  This way we have constructed a bijection between planar $\phi^{3}$ diagrams with a distinguished external leg and planar trees with univalent (white) and bivalent (black) vertices fulfilling the additional condition that  the number of white vertices located above a  given white vertex must be no greater than the number of free edges located below. This is illustrated on a simple example on figure \ref{bijection}. Let us note that this bijection is very similar to the one given by P. Di Francesco in this lecture \cite{difrancesco}.  

The previous bijection is also related to the counting of planar diagrams performed by 't Hooft in \cite{counting}. Indeed, the sum of those trees with weight 1 is a geometric series and solves the functional fixed point equation 
\begin{equation}
F(x)=x+gF^{2}(x)+gh\frac{F(x)-F(0)}{x}.\label{planarsd}
\end{equation}
Note that here  we work with planar trees and count as different two trees that differ by an automorphism. This is why we do not divide by the symmetry factor in the geometric series. 
$F$ is the generating function for trivalent planar diagrams with a distinguished external leg
\begin{equation}
F(x)=\sum_{m,n,p}a_{m,n,p}\,g^{m}h^{n}x^{p},
\end{equation}
with $a_{m,n,p}$ the number of such diagrams with $m$ vertices, $n$ loops and $p$ external legs (not including the distinguished one). Because of the preivous bijection.The first term in the planar Schwinger-Dyson equation \eqref{planarsd} generates the black vertices and the second one the white ones. Besides, the substraction of $F(0)$ ensures that there is a  black leaf related to the white vertex (otherwise $F(x)-F(0)$ vanishes) and the division by $x$ removes this leaf since it is connected to the vertex we have added to the Feynman diagram.  Because of the previous bijection between trees and diagrams, $a_{m,n,p}$ also equals the number of connected planar $\phi^{3}$ Feynman diagrams with a distinguished external leg, $m$ vertices, $n$ loops and $p$ external legs in addition to the distinguished one. For example, up to order 3 in $g$, we find
\begin{equation}
F(x)=x\,+\,g(x^{2}+h)+g^{2}(2x^{3}\,+\,3hx)+g^{3}(5x^{4}+10hx^{2}+4h^{2})\,+\,\dots
\end{equation}
If we set $h=0$, we obtain the Catalan numbers $1,2,5,\dots$ that count the number of planar binary trees.

%\bigskip
%\fbox{\sf Equivalent form of the equation}
%\begin{equation}
%F(z)=g+z+g\frac{gF^{2}(g)-zF^{2}(z)}{g-z}
%\end{equation}
%\bigskip

%\bigskip
%\fbox{\sf Solution of the unrenormalized equation}
%\begin{equation}
%F_{0}(z)=\frac{1}{2g}\left(1-\frac{h}{z}\right)\left(1-\sqrt{1-\frac{4gz}{\left(1-\frac{h}{z}\right)^{2}}}\right)
%=\sum_{n}C_{n}\frac{gz^{n+1}}{\left(1-\frac{h}{z}\right)^{2n-1}}
%\end{equation}
%\bigskip
%It may be that there is some simple subsititution $z\rightarrow z+A(\frac{1}{z},g)$ that cancel all the singularities 

\section{Conclusion and outlook}

\renewcommand{\baselinestretch}{4}
In this talk, we have presented, in a hopefully pedagogical way, some perturbative algebraic methods common to non linear equations and effective action computations. It is conveniently summarized by the following table.
\begin{center}
\begin{tabular}{|c|c|} 
\hline
rooted trees&Feynman diagrams\cr
\hline
non linear analysis & perturbative path integrals\cr
\hline
fixed point equation& renormalization group equation\cr
\hline
powers of non linear operators $X^{t}(x)$&background field technique $A^{\gamma}(S)$\cr
\hline
&\cr
$x'=(\mathrm{id}-X)^{-1}(x)=\sum_{t}\frac{X^{t}(x)}{\mathrm{S}_{t}}$&
$S'[\phi]=\log\int[D\chi]\mathrm{e}^{-\frac{1}{2}\chi\cdot A\cdot\chi+S[\phi+\chi]}=
\sum_{\gamma}\frac{A^{\gamma}(S)}{\mathrm{S}_{\gamma}}[\phi]$\cr
&\cr
\hline
composition&successive integrations\cr
\hline
\end{tabular}
\end{center}

\noindent 
The examples presented here are very simple and can also be treated efficiently using other techniques. However, we think that Hopf algebraic techniques may prove to be useful  when dealing with more complicated situations,  in particular in the following subjects.

\begin{itemize} 
\item
\underline{\bf Multiscale renormalization}

In pertubative QFT, we recover a finite theory if we choose a cut-off dependent bare coupling constant and expand in powers of the low energy renormalized coupling constant. Nevertheless, the expansion is not satisfying since renormalons generally leads to a divergent series. For asymptotically free theories, the divergence of the perturbative perturbative series can cured by expanding not in a single renormalized coupling, but in a scale dependent one \cite{constructive}. The corresponding couplings are obtained by performing the renormalization not in a single step but by successive effective action computations. Therefore, we expect our techniques to be of interest in the construction of the effective expansion in the scale dependent coupling. 

\item
\underline{\bf Graph and matroid polynomials}

Graph and matroid polynomials form a venerable subject where convolution products based on Hopf algebras have already proved to useful (see \cite{KRS}). It may be interesting to further develop the methods presented in the derivation  the universality of the Tutte polynomial and its relation to the $q$-state Potts. In particular, this may useful to investigate properties of the recently proposed multivariate generalizations of the Tutte polynomial \cite{sokal}, or other graph polynomials like the Martin and flow polynomials, which seem to have relations with QFT.

\item
\underline{\bf Spin foam models of quantum gravity}

Loop quantum gravity (see the book \cite{rovelli} for a general overview) is a tentative quantum theory of gravity that provides a construction of the Hilbert space of the quantum gravitational field using spin networks. Its path integral counterpart makes use of higher dimensional analogues of Feynman diagrams termed spin foams. The latter define a Hopf algebra, as uncovered by Markopoulou \cite{Fotini}, which may be useful to address the question of  the Wilsonian renormalization of group field theory \cite{oriti}.

\end{itemize}

\noindent

{\bf Acknowledgements}

We are grateful to F. Girelli who collaborated with us on an early stage of this work. We would also like to thank V. Rivasseau and his group for their \emph{constructive} remarks and M. Aguilar for some comments and a useful reference. Finally, let us also thank K. Ebrahimi-Fard, M. Marcolli and W. van Suijlekom, who organized the conference "Combinatorics and Physics" at the MPIM in Bonn in march 2007. Work partially supported by a EU Marie Curie fellowship EIF-025947 QGNC.

\end{document}